\documentclass[%
 aip,
 amsmath,amssymb,
 reprint,%
floatfix]{revtex4-1}

\usepackage{graphicx}
\usepackage{dcolumn}
\usepackage{bm}

\usepackage[utf8]{inputenc}
\usepackage[T1]{fontenc}
\usepackage{mathptmx}
\usepackage{etoolbox}
\usepackage{booktabs}
\usepackage{lipsum}
\usepackage{algpseudocode}
\usepackage{algorithm}

\makeatletter
\def\@email#1#2{%
 \endgroup
 \patchcmd{\titleblock@produce}
  {\frontmatter@RRAPformat}
  {\frontmatter@RRAPformat{\produce@RRAP{*#1\href{mailto:#2}{#2}}}\frontmatter@RRAPformat}
  {}{}
}%
\makeatother
\begin{document}

\preprint{AIP/123-QED}

\title[Phase kurtosis in diffusion NMR]{Analytical phase kurtosis of the constant gradient spin echo}
\author{Teddy X. Cai}
\affiliation{ \textit{Eunice Kennedy Shriver} National Institute of Child Health and Human Development, Bethesda, MD 20817, USA}
 \email{teddy.cai@nih.gov}
 \author{Nathan H. Williamson}
\affiliation{ \textit{Eunice Kennedy Shriver} National Institute of Child Health and Human Development, Bethesda, MD 20817, USA}
\affiliation{ Military Traumatic Brain Injury Initiative (MTBI\textsuperscript{2}), Bethesda, MD 20814, USA}
\affiliation{ Uniformed Services University of the Health Sciences (USU), Bethesda, Maryland 20814, USA}
\affiliation{ The Henry M. Jackson Foundation for the Advancement of Military Medicine Inc. (HJF), Bethesda, Maryland 20817, USA}
\author{Peter J. Basser}%
 \email{basserp@mail.nih.gov}
\affiliation{ \textit{Eunice Kennedy Shriver} National Institute of Child Health and Human Development, Bethesda, MD 20817, USA}

\date{\today}

\begin{abstract}
The Gaussian phase approximation (GPA) underlies many standard diffusion magnetic resonance (MR) signal models, yet its validity is rarely scrutinized. Here, we assess the validity of the GPA by analytically deriving the excess phase kurtosis $\kappa_4/\kappa_2^2$, where $\kappa_n$ is the $n^{\text{th}}$ cumulant of the accumulated phase distribution due to motion. We consider the signal behavior of the spin echo with constant gradient amplitude $g$ and echo time $T$ in several one-dimensional model systems: (1) a stationary Poisson pore-hopping model with uniform pore spacing $\Delta x$ and mean inter-hop time $\tau_{\text{hop}}$; (2) a trapped-release model in which spin isochromats are initially immobilized and then released with diffusivity $D$ following an exponentially-distributed release time, $\tau_{\text{rel}}$; and (3) restricted diffusion in a domain of length $L$. To our knowledge, this is among the first systematic analytical treatments of spin echo phase kurtosis without assuming Gaussian compartments or infinitesimally short gradient pulses. In the pore-hopping system, $\kappa_4/\kappa^2_2 = (9/5)\tau_{\text{hop}}/T$, inversely proportional to the mean hop number, $T/\tau_{\text{hop}}$. In the trapped-release system, $\kappa_4/\kappa_2^2$ is positive and decreases roughly log-linearly with $T/\langle \tau_{\text{rel}} \rangle$, where $\langle \tau_{\text{rel}} \rangle$ is the average release time. For restriction, $\kappa_4/\kappa_2^2$ vanishes at small and large $L/\sqrt{DT}$, but has complicated intermediate behavior. There is a negative peak at $L/\sqrt{DT} \approx 1.2$ and a positive peak at $L/\sqrt{DT} \approx 4.4$. Monte Carlo simulations are included to validate the analytical findings. Overall, we find that the GPA does not generally hold for these systems under moderate experimental conditions, i.e., $T= 10\;\mathrm{ms}$, $g \approx 0.2 - 0.6\;\mathrm{T/m}$. These results suggest that signal models reliant on the GPA must be carefully examined, particularly for strong, extended gradients and in media with complex kinetics, morphology, and/or microstructure. 
\end{abstract}

\maketitle


\section{Introduction}
Diffusion magnetic resonance (MR) is a powerful means to non-invasively probe microstructure. \cite{Novikov2018b, Grebenkov2007} Fundamentally, such methods measure the amplitude of an echo signal following a refocused magnetic field gradient waveform, or encoding. The signal depends on the shape of the encoding as well as the mobility of spin isochromats during said encoding. In turn, the mobility depends on the microstructure. Relating the signal to microstructural features is generally an ill-posed problem, however, as the relationship between microstructure and signal can be complicated even in seemingly simple cases such as restricted diffusion (i.e., reflecting boundaries). \cite{Robertson1966, Tanner1968, Neuman1974, Grebenkov2007} To make this problem tractable, virtually all diffusion MR signal models adopt simplifying assumptions. \cite{Callaghan1993, Grebenkov2007, Novikov2018a, Novikov2018b}

One such assumption is the Gaussian phase approximation (GPA). It posits that the probability distribution of phase shifts $P(\phi)$ accrued by spin isochromats has negligible higher-order cumulants, such that the signal is determined primarily by the phase variance. The GPA was formalized by Neuman \cite{Neuman1974} in the context of restricted diffusion, though the assumption itself dates to Carr \& Purcell \cite{Carr1954}. Its use is widespread. The GPA is implied in models that assume the signal depends only on a second-order motion parameter such as an apparent diffusion coefficient \cite{Woessner1961, Grebenkov2010} or tensor. \cite{Basser1994} It is explicitly invoked to describe signal in the ``motional averaging'' regime, \cite{Neuman1974, Hurlimann1995} which forms the basis of various multi-compartment signal models. \cite{Novikov2018b} The GPA is also invoked in temporal diffusion spectroscopy (TDS), \cite{Parsons2005, Stepisnik1981, Stepisnik1993} which uses periodically-modulated \cite{Callaghan1995} gradients to probe the spectrum of the velocity autocorrelation function (see also refs. \citenum{Cai2025a, Cai2025b} for related methods). Given its ubiquity, the GPA is a keystone of diffusion microstructural MR.

The GPA, however, is known to break down in certain cases and regimes. A striking example is the ``localization'' regime \cite{Stoller1991, Hurlimann1995, deSwiet1996,  Moutal2020} --- characterized by strong, extended gradients and restriction --- in which the signal $S$ decays asymptotically with gradient amplitude $g$ as $-\ln S \propto g^{2/3}$, in contrast to the Gaussian prediction of $-\ln S \propto g^2$. The GPA can of course be violated under less extreme conditions. As pointed out by Neuman, \cite{Neuman1974} a mixture of Gaussian $P(\phi)$ with distinct variances will, in general, yield a non-Gaussian composite. The GPA can thus be violated for multiple Gaussian compartments, as assumed in bi- \cite{leBihan1991} and multi-exponential signal models. \cite{Yablonskiy2003} The ``diffusion-diffraction'' patterns first observed by Callaghan \cite{Callaghan1991} also clearly contradict the GPA. Finally, various numerical studies have shown that the GPA does not always hold for restricted diffusion, with reported signal deviations of up to $\approx 30\%$. \cite{Balinov1993, Blees1994, Sukstanskii2002, Topgaard2025} Together, these studies suggest that there may be an incongruity in how often the GPA is applied and how often it actually holds.

Understanding when, how, and to what extent the GPA is violated is essential for accurate signal modeling. A natural diagnostic for the GPA is the excess phase kurtosis: $\kappa_4/\kappa_2^2$, where $\kappa_n$ is the $n^{\text{th}}$ cumulant of $P(\phi)$. While the concept of kurtosis has been explored before in diffusion MR, prior research was concerned almost exclusively with the kurtosis of \emph{displacement} distributions, as in the ``diffusional kurtosis'' model \cite{Jensen2005, Jensen2010} (see ref. \citenum{Marrale2015} for review). Displacement and phase, however, are not generally bijective. This holds only in the case of infinitesimally short gradient applications, wherein the phase is simply proportional to displacement. For a realistic, i.e., a finite gradient waveform, phase will accumulate over a \emph{trajectory}. Thus, for experiments with appreciable gradient duration, the displacement kurtosis on its own may not be a reliable diagnostic for the GPA. The excess phase kurtosis, on the other hand, is a direct assessment thereof.

In this manuscript, we derive analytical expressions for $\kappa_4/\kappa_2^2$ in three paradigmatic model systems: (1) Poisson pore-hopping; (2) initial trapping with an exponentially-distributed release time; and (3) restricted diffusion. For clarity and brevity, we consider one-dimensional (1D) diffusion and one experiment: the spin echo \cite{Hahn1950} with constant or static gradients of amplitude $g$. Data from Monte Carlo (MC) simulations are included to validate our findings. To our knowledge, this is one of the only rigorous analytical treatments of higher-order phase statistics in echo signals for non-trivial systems without assuming Gaussian compartments \cite{Fieremans2010, Ning2018} and/or narrow gradient pulses. \cite{Jensen2005, Frhlich2006} Indeed, we are aware of just one other work by Grebenkov \cite{Grebenkov2007b} that explicitly solves for the phase moments in 1D restriction for finite duration gradients, but it does so for a cosine-modulated gradient echo instead.

Our findings suggest the GPA does not generally hold for the constant gradient spin echo (CGSE) in 1D. The Gaussian scaling of $-\ln S \propto g^2$ breaks down and $\kappa_4/\kappa_2^2$ is appreciable for moderate $g \approx 0.1 - 0.6\;\mathrm{T/m}$ and $T \approx 10 \;\mathrm{ms}$. We note that these parameters are readily achieved on pre-clinical diffusion MR scanners and state-of-the-art clinical scanners with high-performance gradients. \cite{Huang2021} These findings suggest that second-order motion statistics may not always suffice to predict the signal, even in simple systems with a single governing time- or lengthscale. That said, we identify trends in the excess phase kurtosis for various systems and clarify the conditions under which the GPA remains valid. 

\section{Basic theory}
\subsection{Cumulant expansion of the phase}
The echo phase $\phi$ resulting from some 1D displacement trajectory of a spin isochromat, or ``spin'' for short,
\begin{equation}
r(t)  = x(t) - x(0),
\end{equation}
is given by
\begin{equation}\label{eq: phase general}
    \phi = \int_0^{T} G(t) r(t) \,dt,
\end{equation}
where $G(t) = \gamma g(t)$ denotes the effective gradient waveform $g(t)$ scaled by the gyromagnetic ratio $\gamma$, and $T$ is the time of echo formation relative to the start of the encoding at $t = 0$ (i.e., excitation). We consider proton diffusion, with $\gamma \approx 2.675222\times 10^{8} \;\mathrm{T^{-1}s^{-1}}$. The relaxation-normalized echo signal $S$ is the expectation
\begin{equation}
    S = \int e^{\mathrm{i}\phi} P(\phi) d\phi = \langle e^{\mathrm{i}\phi} \rangle,
\end{equation}
where again $P(\phi)$ is the spin phase probability distribution, and $\langle\rangle$ denotes averaging over the spin ensemble. We ignore relaxation effects throughout, and focus solely on the signal attenuation due to spin motion. Taking the natural logarithm, the right-hand-side becomes the cumulant generating function (CGF), which can be expanded as \cite{Stepisnik1981}
\begin{equation}\label{eq: phase cumulant expansion}
    \ln S = \sum_{n=1}^{\infty}\frac{\mathrm{i}^n}{n!} \kappa_n,
\end{equation}
where again $\kappa_n$ is the $n^{\text{th}}$ cumulant of $P(\phi)$. Assuming that spin motion is symmetric --- as it would be for thermally-driven self-diffusion --- the odd moments vanish. In this case, $\kappa_2$ is simply the second phase moment, $\kappa_2 = \langle \phi^2 \rangle$.

To a first approximation, deviations from the GPA and Gaussian signal behavior are captured by the fourth cumulant, $\kappa_4 = \langle \phi^4 \rangle - 3\langle \phi^2\rangle^2$, and one can write 
\begin{equation}\label{eq: phase cumulants}
    \ln S^{(4)} = -\frac{\kappa_2}{2} + \frac{\kappa_4}{24}=-\frac{1}{2}\langle \phi^2 \rangle + \frac{1}{24}\left[\langle \phi^4 \rangle -3\langle \phi^2\rangle^2\right],
\end{equation}
where $S^{(n)}$ denotes the signal approximation to $n^{\text{th}}$ order, with the GPA corresponding to $\ln S^{(2)} = -\langle \phi^2 \rangle/2$. Note that the cumulants are related to the kurtosis by 
\begin{equation}
\text{Kurt}(\phi) = \frac{\langle \phi^4 \rangle}{\langle \phi^2 \rangle^2} = \frac{\kappa_4 + 3\kappa^2_2}{\kappa^2_2},   
\end{equation}
and the excess phase kurtosis is merely
\begin{equation}
    \text{Kurt}(\phi) - 3 = \frac{\kappa_4}{\kappa_2^2}.
\end{equation} 
We report $\kappa_4/\kappa_2^2$ as well as second and fourth-order signal approximations $S^{(2)}$ and $S^{(4)}$ throughout. Hereafter, we use the term ``kurtosis'' to refer to $\kappa_4/\kappa_2^2$.

The phase moments $\langle \phi^2 \rangle$ and $\langle \phi^4 \rangle$ that constitute $\kappa_2$ and $\kappa_4$ can be found via ensemble-average displacement correlators. We denote these as
\begin{equation}
    C(t_1, t_2) = \langle r(t_1)r(t_2) \rangle
\end{equation} for the $2$-point, or $2$-time correlator, and 
\begin{equation}
    C(t_1, t_2, t_3, t_4) = \langle r(t_1)r(t_2)r(t_3)r(t_4)\rangle,
\end{equation}
for the $4$-point correlator. In general, the $n^{\text{th}}$ phase moment is given by an $n$-fold time integral over the $n$-point correlator and $G(t)$. Explicitly,
\begin{equation}\label{eq: phi2}
    \langle \phi^2\rangle = \int_0^T\int_0^T G(t_1)G(t_2)C(t_1,t_2)dt_1dt_2
\end{equation}
and
\begin{equation}\label{eq: phi4}
\begin{aligned}
    \langle \phi^4 \rangle = \iiiint_{[0,T]^4}& dt_1 dt_2 dt_3 dt_4\\ &\times G(t_1) G(t_2) G(t_3) G(t_4) C(t_1,t_2,t_3,t_4).
\end{aligned}
\end{equation}  

In the following sections, we study systems of increasing complexity in which closed-form expressions for $\kappa_2$ and $\kappa_4$ can be found, yielding the signal approximation in Eq. \eqref{eq: phase cumulants}. We consider only the classic experimental paradigm of the CGSE, which has the waveform:
\begin{equation}\label{eq: grad spin echo}
    G(t) = \begin{cases} + \gamma g, &  0\leq t < \tfrac{T}{2} \\
                         - \gamma g, & \tfrac{T}{2} \leq t < T
            \end{cases}\;.
\end{equation}
Extension to the more general pulsed-gradient spin echo (PGSE) experiment \cite{Stejskal1965} is discussed briefly towards the end of the manuscript.

\subsection{Free diffusion}

As a trivial but instructive example, let us first consider free diffusion with self-diffusion coefficient $D$. Only the second cumulant is non-zero. The 2-point correlator is
\begin{equation}\label{eq: 2-point corr free}
    C(t_1, t_2) = 2D\,\min{(t_1, t_2)},
\end{equation}
which is simply the covariance of two Brownian processes. The second phase moment is given according to Eq. \eqref{eq: phi2} by
\begin{equation*}
    \langle \phi^2 \rangle = 4D\int_0^T\int_0^{t_2} G(t_1)G(t_2) t_1 dt_1 dt_2,
\end{equation*}
using the time-symmetry of the correlator. Evaluating the above integral, $I$, over piecewise domains:
\begin{equation*}
    I = 4\gamma^2g^2 T^3 D\begin{cases}
          1/48, & t_1,t_2 \in \left[0, \frac{T}{2}\right] \\
          1/12, & t_1,t_2 \in \left[\frac{T}{2}, T\right] \\
          -1/16, & t_1 \in \left[0, \frac{T}{2}\right], t_2\in\left[\frac{T}{2}, T\right]
    \end{cases}\;,
\end{equation*}
noting that the gradient terms can be handled as $G(t_1)G(t_2) = (\gamma g)^2\,\text{sgn}\left(t_1 - \tfrac{T}{2}\right)\text{sgn}\left(t_2 - \tfrac{T}{2}\right)$, where $\text{sgn}$ is the sign function. Summing and plugging into Eq. \eqref{eq: phase cumulants}, one obtains the well-known result for the spin echo signal \cite{Carr1954} due to free diffusion:
\begin{equation}\label{eq: decay free}
    \ln S = -\frac{\langle \phi^2\rangle}{2}=-\frac{\gamma^2g^2T^3D}{12},
\end{equation}
which is often written as
\begin{equation}
    \ln S = - bD, \quad b = \frac{\gamma^2 g^2 T^3}{12}
\end{equation}
where $b$ is the $b$-value that characterizes the overall diffusion weighting. It is more generally defined as \cite{Stejskal1965}
\begin{equation}\label{eq: b-val}
    b = \int_0^T \left(\int_0^t G(t')dt'\right)^2 dt.
\end{equation}
While this formulation in the $b$-value gives the correct result in the case of free diffusion, we will show in the following sections that the $b$-value is not always a meaningful way to describe the signal behavior.

\section{Poisson pore-hopping model}

The first non-trivial system we consider is a special case of continuous-time random walk: a stationary Poisson pore-hopping process. In particular, we assume that $r(t)$ is given by a Poisson counting process,
\begin{equation}\label{eq: Poisson displacement}
    r(t) = \Delta x \sum_{j = 1}^{H(t)} \zeta_{j},
\end{equation}
where $\Delta x$ is a uniform pore spacing, $\zeta_j$ denotes i.i.d. random variables taking values $\pm 1$ with equal probability, and $H(t)$ is the hop count with probability
\begin{equation}\label{eq: Poisson count}
    P(H(t) = h) = \frac{\left(t/\tau_{\text{hop}}\right)^h}{h!} e^{-t/\tau_{\text{hop}}},
\end{equation}
where $\tau_{\text{hop}}$ is the mean inter-hop time. 

While this system is not expressly physical (at least not for water in biological tissue), it can be viewed as a simplified representation of small, periodic pores separated by weakly permeable barriers such that $\tau_{\text{hop}}$ is much longer than the time it takes for spins to sample or equilibrate within the pore --- small here meaning that phase accrual within the pore can be neglected. The pore-to-pore transmission in said case (i.e., barrier-limited exchange) is well-modeled by a first-order rate process, as above, wherein the rate $1/\tau_{\text{hop}}$ is proportional to the pores' surface-to-volume ratio. Given fast equilibration, the displacement per transmission should be $\approx \Delta x$. 

\subsection{Second moment}
The correlator $C(t_1, t_2)$ is given from Eq. \eqref{eq: Poisson displacement} as
\begin{equation*}
    C(t_1, t_2) = (\Delta x)^2 \sum_{j_1 = 1}^{H(t_1)} \sum_{j_2 = 1}^{H(t_2)} \left\langle \zeta_{j_1} \zeta_{j_2} \right\rangle.
\end{equation*}
The double sum of $\left\langle \zeta_{j_1} \zeta_{j_2} \right\rangle$ is the product of the number of common hops up to $\text{min}(t_1, t_2)$, as it will be $0$ for non-overlapping times. The sum therefore yields $\text{min}(t_1, t_2)/\tau_{\text{hop}}$, noting that $\langle \zeta_j^2 \rangle = 1$. Unsurprisingly, given that this is still fundamentally a random walk process, one obtains the same structure given in Eq. \eqref{eq: 2-point corr free} for free diffusion:
\begin{equation}
    C(t_1, t_2) = \frac{(\Delta x)^2}{\tau_{\text{hop}}} \text{min}(t_1, t_2).
\end{equation}
One can define an effective diffusivity from $\langle r^2(t) \rangle = 2Dt = (\Delta x)^2t/\tau_{\text{hop}}$ of
\begin{equation}\label{eq: Deff poisson}
    D_{\text{eff}} = \frac{(\Delta x)^2}{2\tau_{\text{hop}}}.
\end{equation}
Finding $\langle \phi^2 \rangle$ proceeds in the same manner as the previous subsection, and one obtains
\begin{equation}\label{eq: phi2 poisson}
    \langle \phi^2 \rangle = \frac{\gamma^2 g^2 T^3 D_{\text{eff}}}{6} = \frac{\gamma^2 g^2  T^3 (\Delta x)^2}{12\tau_{\text{hop}}}, 
\end{equation}
analogous to Eq. \eqref{eq: decay free}. 

\subsection{Higher-order cumulants}\label{sec: Poisson higher-order}

For this particular system, it is convenient to find $\kappa_4$ and higher-order cumulants via the CGF, rather than displacement correlators. Recall that the CGF is merely $\ln \langle e^{\mathrm{i}\phi} \rangle$. To begin, consider the conditional, single-hop expectation. For some hop $\zeta_j$ at time $\tau_j$ (assuming that hopping is an instantaneous event), the phase contribution from this hop will be 
\begin{equation}
    \Delta \phi_j =  \zeta_j \Delta x \left(\int_{\tau_j}^T G(t)dt\right),
\end{equation} 
where the integral describes the running accumulation of phase throughout the remainder of the encoding. Let us call this integral: $F(\tau_j) = \int_{\tau_j}^T G(t) dt$. The echo phase $\phi$ is given by the sum of all $\Delta \phi_j$ up to $H(T)$. It is obvious that for a given $H(T) = h$, i.e., further conditioning on hop count, that
\begin{equation*}
    \mathbb{E}_{\tau, \zeta}\left[e^{\mathrm{i}\sum_{j=1}^{h} \Delta \phi_j} \right] = \left(\mathbb{E}_{\tau, \zeta}\left[e^{\mathrm{i}\Delta \phi_j} \right]\right)^h,
\end{equation*}
where $\mathbb{E}_{\tau, \zeta}$ denotes the conditional expectation on event time and $\zeta$. Using Eq. \eqref{eq: Poisson count}, one can relate the echo signal $S = \langle e^{\mathrm{i}\phi} \rangle$ to the conditional expectation above as
\begin{equation*}
    \langle e^{\mathrm{i}\phi} \rangle = \sum_{n = 0}^{\infty}\frac{\left(T/\tau_{\text{hop}}\right)^h}{h!} e^{-T/\tau_{\text{hop}}}\left(\mathbb{E}_{\tau, \zeta}\left[e^{\mathrm{i}\Delta \phi_j} \right]\right)^h.
\end{equation*}
Taking the log and rearranging, one obtains a preliminary form of the log signal:
\begin{equation}\label{eq: CGF preliminary}
    \ln S = \frac{T}{\tau_{\text{hop}}}\left[\left(\mathbb{E}_{\tau, \zeta}\left[e^{\mathrm{i}\Delta \phi_j} \right]\right) - 1\right].
\end{equation}

Let us now take a closer look at the single-hop expectation. It can be written as a time-average over $\tau_j$ (which is sampled uniformly on $[0, T]$) and a second average over $\zeta$:
\begin{equation*}
    \left(\mathbb{E}_{\tau, \zeta}\left[e^{\mathrm{i}\Delta \phi_j} \right]\right) = \frac{1}{T}\int_0^T \mathbb{E}_{\zeta}\left[ e^{\mathrm{i}F(\tau_j)\Delta x \zeta }\right] d\tau_j,
\end{equation*}
where $\mathbb{E}_\zeta$ denotes averaging over the possible values of $\zeta$. As $\zeta$ is equally likely to be $\pm 1$, the average yields
\begin{equation*}
    \mathbb{E}_{\zeta}\left[ e^{\mathrm{i} \zeta F(\tau_j)\Delta x}\right] = \frac{1}{2}\left(e^{\mathrm{i}\Delta x F(\tau_j)} + e^{-\mathrm{i}\Delta x F(\tau_j)}\right) = \cos( F(\tau_j)\Delta x).
\end{equation*} 
Plugging back into Eq. \eqref{eq: CGF preliminary}:
\begin{equation*}
    \ln S = \frac{1}{\tau_\text{hop}} \int_0^T \left[\cos\left( F(\tau_j)\Delta x\right) - 1\right] d\tau_j.
\end{equation*}
From the gradient waveform in Eq. \eqref{eq: grad spin echo}, one finds that $F(\tau_j)$ for this experiment is simply
\begin{equation}\label{eq: F(tau_j)}
    F(\tau_j) = \int_{\tau_j}^T G(t) dt = -\gamma g\; \text{min}(\tau_j, T - \tau_j).
\end{equation}
The previous integral for $\ln S$ can thus be rewritten as
\begin{equation*}
    \ln S = \frac{2}{\tau_\text{hop}}\int_0^{T/2} \left[\cos(\gamma g \tau_j \Delta x) - 1\right]  d\tau_j,
\end{equation*}
using the fact that cosine is even. This evaluates to
\begin{equation}\label{eq: poisson signal}
    \ln S = \frac{2}{\tau_{\text{hop}}}\left[\frac{\sin(\gamma g \Delta x (T /2))}{\gamma g \Delta x }- \frac{T}{2}\right],
\end{equation}
which is an exact signal expression for this particular system and experiment.

One can recover the cumulant representation of the signal by Taylor expansion. Let $u = \gamma g \Delta x (T/2)$ such that the above becomes
\begin{equation*}
    \ln S = \frac{T}{\tau_{\text{hop}}}\left(\frac{\sin(u)}{u} - 1\right).
\end{equation*}
Using the expansion, $\sin(u)/u - 1 = -x^2/3! + x^4/5!\hdots$, and rearranging, one obtains
\begin{equation}
    \ln S = \frac{1}{\tau_{\text{hop}}}\sum_{m = 1}^\infty (-1)^m \frac{(\gamma g \Delta x)^{2m} T^{2m + 1}}{2^{2m}(2m + 1)!}
\end{equation}
By comparison to Eq. \eqref{eq: phase cumulant expansion}, one then finds that the cumulants are given by
\begin{equation}\label{eq: nth cumulant poisson}
    \kappa_n = \frac{1}{\tau_{\text{hop}}} \frac{(\gamma g \Delta x)^n \,T^{n + 1}}{2^n (n+1)}, \quad \text{for}\;\text{even}\; n\;\text{only},
\end{equation}
and $0$ otherwise. This is consistent with the result for $\kappa_2 = \gamma^2g^2T^3(\Delta x)^2/(12\tau_{\text{hop}})$ given in Eq. \eqref{eq: phi2 poisson}. 

The excess phase kurtosis is therefore
\begin{equation}\label{eq: poisson kurt}
    \frac{\kappa_4}{\kappa_2^2} = \frac{9}{5} \frac{\tau_{\text{hop}}}{T}.
\end{equation}
This is strictly positive (i.e., $P(\phi)$ is leptokurtic), and we note that it is inversely proportional to $T/\tau_{\text{hop}}$, or simply the mean number of hops during the encoding. Thus, the GPA can be said to hold when there are many hops: $T \gg \tau_{\text{hop}}$. Intuitively, for few hops, $P(\phi)$ will be peaked at $\phi = 0$, yielding a positive kurtosis. Gaussianity is then restored for a sufficient number of hops by the central limit theorem (CLT). We expect that the proportionality between $\kappa_4/\kappa_2^2$ and the mean hop number should hold for single echo experiments, while the specific $9/5$ pre-factor emerges for just for the CGSE, see below. 

It is important to note that this result implies that, from a distributional point-of-view, deviation from a Gaussian $P(\phi)$ is not dependent on $g$. From a signal point-of-view, however, non-Gaussian effects will be more pronounced for greater $g$ and signal dephasing. This is because the difference between $\ln S^{(2)}$ and $\ln S^{(4)}$, i.e., $\kappa_4/24$ --- see again Eq. \eqref{eq: phase cumulants} --- scales with the raw cumulant $\kappa_4 \propto g^4$ rather than the kurtosis.  

Returning to the pre-factor, let us also consider the case of instantaneous gradients at times $t = 0,\,T/2$, or a PGSE experiment with narrow pulses. In this case, $F(\tau_j)$ in Eq. \eqref{eq: F(tau_j)} becomes $-\gamma g$ for $0 \le \tau_j < T/2$, and $0$ otherwise. The derivation proceeds in the same manner, and one obtains $\kappa_4/\kappa_2^2 = 2(\tau_{\text{hop}})/T$, a slightly larger pre-factor than $9/5$. The CGSE case thus exhibits less kurtosis. This is because the effect of a hop is spread over the encoding for CGSE, rather than occurring as a sharp transition as in PGSE.  

\subsection{Signal behavior}

\begin{figure}
    \centering
    \includegraphics{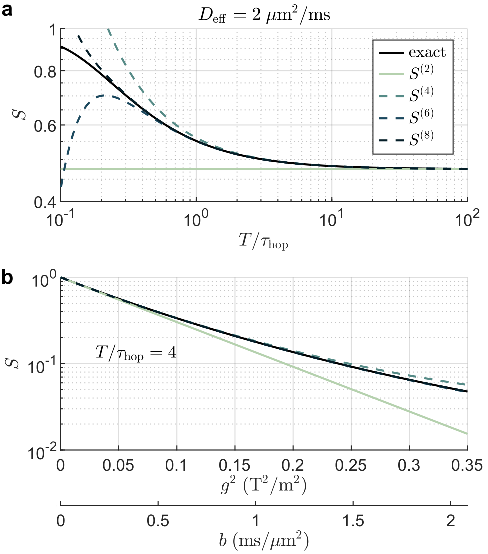}
    \caption{Exact signal (solid, black) and $n^{\text{th}}$-order approximations for the Poisson pore-hopping model. (a) Signal vs. $T/\tau_{\text{hop}}$, where $T = 10\;\mathrm{ms}$, $g = 0.25\;\mathrm{T/m}$ were fixed, and $\tau_{\text{hop}}$ was varied between $0.01 - 100\;\mathrm{ms}$, linearly spaced by $0.01\;\mathrm{ms}$. The overall mobility was also fixed at $D_{\text{eff}} = 2\;\mathrm{\mu m^2/ms}$, with the spacing scaled according to Eq. \eqref{eq: Deff poisson}: $\Delta x = \sqrt{2D_{\text{eff}}\tau_{\text{hop}}}$. The GPA corresponds to $S^{(2)}$ (solid, light green). Higher-order approximations (dashed) are also shown. (b) Signal vs. $g^2$, with $g$ varied linearly between $0 - 0.6\;\mathrm{T/m}$ by $0.01$, and $T/\tau_{\text{hop}} = 4$ fixed. Other parameters were the same as part (a), and the same legend applies. A secondary axis with $b = \gamma^2g^2T^3/12$ is included for reference.}
    \label{fig: poisson signal}
\end{figure}

In Fig. \ref{fig: poisson signal}, the exact signal in Eq. \eqref{eq: poisson signal} is compared to the $n^{\text{th}}$-order approximations $S^{(n)}$. In Fig. \ref{fig: poisson signal}a, the signal is plotted vs. the mean hop number $T/\tau_{\text{hop}}$, with fixed $T = 10 \;\mathrm{ms}$ and $g = 0.25 \;\mathrm{T/m}$, varying $\tau_{\text{hop}}$. From Eq. \eqref{eq: b-val}, this corresponds to a conventional $b$-value of $b \approx 0.37\;\mathrm{ms/\mu m^2}$. To facilitate a meaningful comparison, the effective diffusivity was fixed to be $D_{\text{eff}} = 2 \;\mathrm{\mu m^2/ms}$ such that the hop spacing $\Delta x$ scales with $\tau_{\text{hop}}$ as $\Delta x = \sqrt{2D_{\text{eff}}\tau_\text{hop}}$. That is, the overall mobility or mean-squared displacement was kept the same, while the mean number of hops needed to achieve said displacement was varied. Since $D_{\text{eff}}$ was fixed, $\kappa_2$, which contains only second-order displacement information, will not vary, and $S^{(2)}$ is expectedly a flat line --- see again Eqs. \eqref{eq: Deff poisson} and \eqref{eq: nth cumulant poisson}.

It can be seen that $S^{(2)}$ is a good approximation for $S$ as the mean hop number increases ($T/\tau_{\text{hop}} \gtrsim 10$), but fails for small $T/\tau_{\text{hop}}$. The higher-order approximations converge to $S$, with $S^{(4)}$ remains a good approximation in the range of $T/\tau_{\text{hop}} \sim 1 - 10$. The behavior of these higher-order approximations, however, is not physical over the considered domain. That is, $S^{(n)}$ may exceed 1 and/or exhibit non-monotonic behavior, particularly at small $T/\tau_{\text{hop}}$.    

Related to this behavior, the Pawula theorem \cite{Pawula1967a, Pawula1967b} states that a truncated approximation in cumulants such as $S^{(4)}$ does not correspond to a valid $P(\phi)$ (i.e., one that is positive-definite and normalizable). This property is guaranteed only for second-order and infinite-order representations. Thus, it is not surprising that the approximations $S^{(n>2)}$ exhibit non-physical behavior at high $g$, akin to a radius of convergence. \cite{Frhlich2006} Despite this, higher-order approximations remain useful, and are more accurate than $S^{(2)}$ in certain regimes, as seen here.

In Fig. \ref{fig: poisson signal}b, the signal approximations are plotted vs. $g^2$ for a fixed value of $T/\tau_{\text{hop}} = 4$, keeping other parameters the same as Fig. \ref{fig: poisson signal}a. The same legend applies. We choose $T/\tau_{\text{hop}} = 4$ because it yields a comparable $\kappa_4/\kappa_2^2 \approx 0.45$ to the maximal kurtosis seen in the restricted system considered later in this manuscript. On this axis, the second-order approximation $S^{(2)}$ decays linearly. Higher-order approximations and the exact signal decay sub-linearly, consistent with positive $\kappa_4/\kappa_2^2$. The approximations $S^{(6)}$ and $S^{(8)}$ are close to the exact signal, while $S^{(4)}$ is accurate until $S \lesssim 0.1$, and $S^{(2)}$ until $S \lesssim 0.7$.


\section{Trapped-release model with exponentially-distributed release time}\label{sec: exp trap}

We next consider a system in which particles are initially trapped at a point ($x =0$), and released after a probabilistic time $\tau_{\text{rel}}$ drawn from an exponential distribution:
\begin{equation}
    P(\tau_{\text{rel}}) = ke^{-k\tau_{\text{rel}}},
\end{equation}
where $k = 1/\langle \tau_{\text{rel}}\rangle$ is the inverse of the average release time. Following release, spins are assumed to diffuse freely with $D$, preserving symmetry in $P(\phi)$. Note that re-entry or re-trapping is ignored. This system extends the previous case of Poisson pore-hopping by considering what is effectively a small pore that interfaces not with self-similar pores, but with a free compartment. This system can be viewed as a simplified representation of many such pores in the case of strong diffusion weighting. If the diffusion weighting is strong enough to fully dephase spin trajectories that begin in the free compartment, then ignoring both re-entry and spins that are initially free may be justified, i.e., all such signal vanishes, and the behavior of the remaining signal will be dominated by release dynamics of trapped spins.

\subsection{Second moment}
Let us first find the 2-point correlator. The correlator for a given release time $\tau_{\text{rel}}$ (i.e., conditioned on $\tau_{\text{rel}}$) is 
\begin{equation} \label{eq: exp trap conditional}
    C(t_1, t_2 \mid \tau) = 2D\left[\text{min}(t_1,t_2)-\tau_{\text{rel}}\right]\, \theta(\text{min}(t_1,t_2) - \tau_{\text{rel}}),
\end{equation}
where $\theta(t)$ is the Heaviside step function. The unconditioned correlator is given by the average:
\begin{equation*}
    C(t_1,t_2) = 2Dk\int_0^{\text{min}(t_1,t_2)} \left[\text{min}(t_1,t_2)-\tau_{\text{rel}}\right]  e^{-k\tau} d\tau_{\text{rel}}.
\end{equation*}
Let $u = \text{min}(t_1,t_2) -\tau_{\text{rel}}$ and note that $\int_0^a ue^{ku} du = ae^{ka}/k + (1 -e^{ka})/k^2$. One obtains  
\begin{equation}\label{eq: exp trap 2-point}
    C(t_1, t_2) = 2D\left[\text{min}(t_1,t_2) - \frac{1 - e^{-k\,\text{min}(t_1,t_2)}}{k}\right],
\end{equation}
and thus,
\begin{equation*}
    \langle \phi^2 \rangle = 4D\int_0^T\int_0^{t_2} G(t_1)G(t_2)\left( t_1 - \frac{1-e^{-kt_1}}{k}\right)dt_1dt_2,
\end{equation*}
making use of time symmetry. Note that the $t_1$ term in the above yields the same result given previously in Eq. \eqref{eq: decay free} for free diffusion. Evaluating the remaining terms in the integral, $I = -4Dk^{-1}\int_0^T\int_0^{t_2}G(t_1)G(t_2)(1-e^{-kt_1})dt_1dt_2$, piecewise:
\begin{multline*}
    I = 4\gamma^2g^2 (T/\alpha)^3 D \times \\ \begin{cases}
        - \frac{\alpha^2}{8} +\frac{\alpha}{2} - (1 - e^{-\alpha/2})  , & t_1, t_2 \in \left[0, \frac{T}{2}\right] \\
         - \frac{\alpha^2}{8} + e^{-\alpha/2}(\frac{\alpha}{2}  - 1) + e^{-\alpha} , & t_1, t_2 \in \left[\frac{T}{2}, T\right] \\
        \frac{\alpha^2}{4} + \frac{\alpha}{2}(e^{-\alpha/2}-1), & t_1 \in \left[0, \frac{T}{2}\right], t_2\in\left[\frac{T}{2}, T\right],
    \end{cases}
\end{multline*}
where we define the dimensionless quantity, 
\begin{equation}
    \alpha \equiv kT = T/\langle\tau_{\text{rel}}\rangle
\end{equation}
as the ratio of $T$ and the mean release time $\langle \tau_{\text{rel}} \rangle$. Note that the unreleased spin fraction is given by $1 - e^{-\alpha}$. Summing and adding back the free diffusion result:
\begin{equation}\label{eq: exp trap phi2}
    \langle \phi^2\rangle = \gamma^2g^2 T^3 D\left(\frac{1}{6} - \frac{4}{\alpha^3}\left[1 - e^{-\alpha} - \alpha e^{-\alpha/2} \right]\right).
\end{equation}
Note that the square bracketed term $\rightarrow 0$ at $\lim_{\alpha\rightarrow \infty}$ or $\lim_{k\rightarrow\infty}$, which corresponds to instantaneous release. This leaves the free diffusion term, as expected. At $\lim_{\alpha\rightarrow 0}$, corresponding to no release, this term approaches $ -1/6$. This can be seen by expanding $e^{-\alpha} =\sum_{m=0}^{\infty}(-1)^m\alpha^m/m!$ to leading order. Thus, $\lim_{\alpha\rightarrow0}\langle \phi^2\rangle = 0$, consistent with spins being trapped.   

\subsection{Fourth moment}

To find $\langle \phi^4\rangle$, we will consider the 4-point correlator. When conditioned on $\tau_{\text{rel}}$, the correlator can be decomposed as
\begin{equation}
    C(t_1,t_2,t_3, t_4 \mid \tau_{\text{rel}}) = \sum_{[a,b],[c,d]\in W} C(t_a, t_b \mid \tau_{\text{rel}}) \,C(t_c, t_d \mid \tau_{\text{rel}}),
\end{equation}
where 
\begin{equation*}
    \mathcal{W} = \{[1,2],[3,4];\, [1,3],[2,4];\, [1,4],[2,3]\}
\end{equation*}
is the set of pairings of 4 indices into 2 disjoint pairs, also known as Wick \cite{Wick1950} pairs. Isserlis's theorem \cite{Isserlis1918} allows the $4$-point correlator to be expressed in this way, and each Wick pair consists of a product of two instances of Eq. \eqref{eq: exp trap conditional}. Note that Isserlis's theorem applies specifically for the conditioned correlator as it is either $0$, or zero-mean and jointly Gaussian, satisfying the conditions for the theorem. This decomposition simplifies the problem into a sum of combinations of 2-point correlators. Averaging over $\tau_{\text{rel}}$ yields the unconditional 4-point correlator:
\begin{multline}
    C(t_1,t_2,t_3, t_4) = 4D^2k\sum_{[a,b],[c,d]\in W} \\ \int_0^{t_<} \left[\text{min}(t_a,t_b) - \tau_{\text{rel}}\right]\left[\text{min}(t_c,t_d) -\tau_{\text{rel}}\right]e^{-k\tau_{\text{rel}}}d\tau_{\text{rel}},
\end{multline}
where for convenience we denote 
\begin{equation*}
t_< \equiv \text{min}(t_1,t_2,t_3,t_4).
\end{equation*}
For a given Wick pair, the integral above solves as 
\begin{widetext}
\begin{equation}\label{eq: exp trap wick pair}
\begin{aligned}
    I_{[a,b],[c,d]} = \text{min}(t_a,t_b)\,\text{min}(t_c,t_d)\left(\frac{1-e^{-kt_<}}{k}\right) -\left[\text{min}(t_a, t_b) + \text{min}(t_c,t_d)\right] \left(\frac{1 - e^{-kt_<}\left[1+kt_<\right]}{k^2}\right)  - \frac{2 + e^{-kt_<}\left(k^2t^2_<+2kt_<+2\right)}{k^3},
\end{aligned}
\end{equation}
obtained by expanding and using the previous result in Eq. \eqref{eq: exp trap 2-point}, and noting that $\int_0^a u^2 e^{-ku} du$ has the form of the last term. The fourth moment is given by substituting into Eq. \eqref{eq: phi4}:
\begin{equation*}
\begin{aligned}
    \langle \phi^4\rangle =  4D^2k\iiiint_{[0,T]^4} G(t_1)G(t_2)G(t_3)G(t_4) \sum_{[a,b],[c,d]\in \mathcal{W}} I_{[a,b],[c,d]} \;dt_1 dt_2 dt_3 dt_4.
\end{aligned}
\end{equation*}

We can take the same general approach of splitting into sub-domains and evaluating piecewise, but the process is more involved as the domain is a 4D hypercube. Details on a symbolic programming approach are presented in Appendix \ref{appx: exp trap integrals}. In short, each of the four times is assigned to either $\left[0,\tfrac{T}{2}\right]$ or $\left[\tfrac{T}{2}, T\right]$, and the $4!=24$ time orderings are permuted to cover the hypercube per Wick pair, with the integrand adjusted as appropriate. The parity of the assignment determines the sign of $G(t_1)G(t_2)G(t_3)G(t_4)$. Altogether, one obtains:
\begin{equation}\label{eq: exp trap phi4}
    \langle \phi^4 \rangle = \frac{\gamma^4g^4 T^6 D^2}{12\alpha^6}\left(\alpha^6 + 24\alpha^4e^{-\alpha/2} -\alpha^3 \left[48 + 432\,e^{-\alpha/2}\right] + 11,520\, \left[1 - e^{-\alpha}-\alpha e^{-\alpha/2}\right] \right).
\end{equation}
\end{widetext}
Note that $\lim_{\alpha\rightarrow0}\langle \phi^4\rangle=0$, as expected for trapped spins. Note too that the leading order term in $\langle \phi^4 \rangle$ is $=\gamma^4g^4D^2T^6/12$, which is the same as the leading free diffusion term in $3\langle \phi^2 \rangle^2 \simeq 3(\gamma^2g^2DT^3/6)^2$. This implies $ \lim_{\alpha\rightarrow\infty}\langle \phi^4\rangle/\langle\phi^2\rangle^2 = 3$, and thus $\lim_{\alpha\rightarrow\infty}\kappa_4/\kappa_2^2 = 0$, as expected for instantaneous release and thereby Gaussian diffusion. It can also be shown that $\langle \phi^4\rangle/\langle\phi^2\rangle^2 > 3$ for $\alpha > 0$ given some fixed $T$. Thus, $\kappa_4/\kappa_2^2>0$ and $P(\phi)$ is leptokurtic, consistent with heavy tails and a sharp central peak at $\phi = 0$, as expected for a model of escape events. 

\begin{figure}
    \centering
    \includegraphics{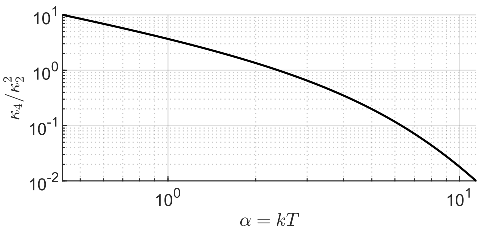}
    \caption{$\kappa_4/\kappa_2^2$ vs. $\alpha = kT$ for the trapped-release model with mean release time $1/k$. This behavior does not depend on $D$ or $g$. }
    \label{fig: trapping release kurt}
\end{figure}

Similar to the Poisson pore-hopping system, $\kappa_4/\kappa_2^2$ cancels $g$ and $D$ such that deviation from Gaussianity in $P(\phi)$ is solely a function of $\alpha = kT$. In Fig. \ref{fig: trapping release kurt}, $\kappa_4/\kappa_2^2$ is plotted vs. $\alpha$. The kurtosis monotonically decreases, at first approximately log-linearly, and then at a faster rate as $\alpha \gtrsim 5$. This regime corresponds to nearly all spins being released at some point during the encoding --- note, $1 - e^{-5} \approx 0.99$.

\subsection{Predicted signal behavior}

\begin{figure}
    \centering
    \includegraphics{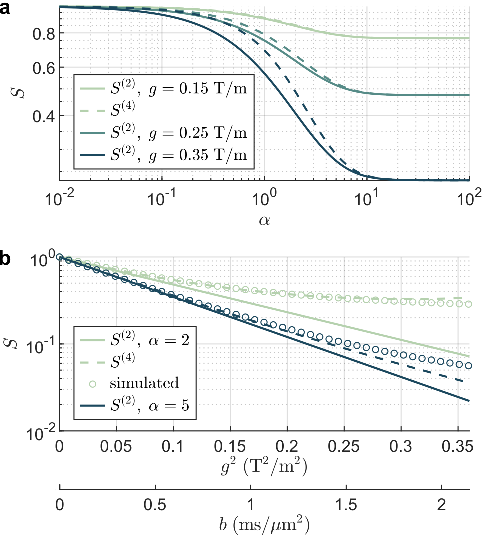}
    \caption{Signal approximations and simulated signal for the trapped-release model. (a) $S^{(2)}$ (solid) and $S^{(4)}$ (dashed) vs. $\alpha = kT$, with $\alpha$ varied log-linearly between $10^{-2} - 10^2$ at $200$ points. The echo time $T = 10\;\mathrm{ms}$ and $D = 2\;\mathrm{\mu m^2/ms}$ were fixed, while $g =$ [0.15, 0.25, 0.35] $\mathrm{T/m}$ was varied (colors). (b) $S^{(2)}$ (solid), $S^{(4)}$ (dashed), and MC simulated signal (circles) vs. $g^2$ for fixed values of $\alpha =[2,\,5]$ (colors), with $g$ varied linearly between $0-0.6$ by $0.01\;\mathrm{T/m}$. Other parameters were kept the same as part (a). A secondary axis with $b = \gamma^2 g^2 T^3/12$ is included for reference. Note that $S^{(2)}$ scales with $g^2$ while $S^{(4)}$ scales sub-linearly.}
    \label{fig: exp trap release analyt}
\end{figure}

In Fig. \ref{fig: exp trap release analyt}a, $S^{(2)}$ and $S^{(4)}$ are plotted vs. $\alpha$ for fixed $T = 10\;\mathrm{ms}$, $D = 2\;\mathrm{\mu m^2/ms}$, and various $g =$ [0.15, 0.25, 0.35] $\mathrm{T/m}$. This corresponds to $b$-values of $b \approx $ [0.13, 0.37, 0.73] $\mathrm{ms/\mu m^2}$. As discussed, $\kappa_4$ is positive such that $S^{(4)}$ exhibits \emph{less} dephasing than $S^{(2)}$. The deviation between $S^{(4)}$ and $S^{(2)}$ vanishes at the limits of $\alpha \gg 1$ or $\alpha \ll 1$, and the GPA is accurate when $\alpha \gtrsim 5$, as expected from Fig. \ref{fig: trapping release kurt}. It is also worth pointing out that the greatest signal deviation between $\ln S^{(2)}$ and $\ln S^{(4)}$ occurs at about $\alpha \approx 1.3$, which corresponds to a maximum in $\kappa_4$ (data not shown).

 In Fig. \ref{fig: exp trap release analyt}b, the signal approximations are instead plotted vs. $g^2$ for fixed values of $\alpha =$ [2, 5], keeping $T = 10\;\mathrm{ms}$ and $D = 2\;\mathrm{\mu m^2/ms}$ the same. Again, $\ln S^{(2)}$ scales linearly with $g^2$, while $\ln S^{(4)}$ decays sub-linearly.
It can be seen that non-physical behavior emerges in $S^{(4)}$ as $g$ increases --- see the regime $g^2 > 0.3\;\mathrm{T^2/m^2}$ for $\alpha =2$ in Fig. \ref{fig: exp trap release analyt}b, where $S^{(4)}$ begins to increase. This is expected by the Pawula theorem, and indicates that higher-order cumulants $n > 4$ can not be ignored in the regime $g^2 \gtrsim 0.3 \;\mathrm{T^2/m^2}$.


Here, we do not have an exact expression for the signal with which to compare. To validate our results, we have included numerically simulated signals in Fig. \ref{fig: exp trap release analyt}b. These data were generated by MC simulations with time step $2.5\;\mathrm{\mu s}$, step size $0.1 \;\mathrm{\mu m}$ (binary step direction, scaled according to time step and $D$), and $1\times 10^4$ walkers. The simulation was run once per $g$. It can be seen that $S^{(4)}$ agrees closely with the simulated signal for $\alpha = 2$, but diverges somewhat at large $g^2 \gtrsim 0.3\;\mathrm{T^2/m^2}$ for $\alpha = 5$. This deviation lies in a regime in which the signal is largely dephased ($S \lesssim 0.1$). Again, higher-order terms may be needed to accurately describe strong signal dephasing. We note that the simulated $P(\phi)$ agrees with Fig. \ref{fig: trapping release kurt} (data not shown), i.e., the kurtosis itself is correct and the discrepancy in Fig. \ref{fig: exp trap release analyt}b does in fact arise from neglected higher-order cumulants. Similar to Fig. \ref{fig: poisson signal}b, the GPA or $S^{(2)}$ is a reasonably accurate approximation for $S\gtrsim 0.7$.

\section{Restricted diffusion}

Lastly, we consider diffusion restricted to the domain $x\in [0,L]$, i.e., with Neumann boundary conditions. We assume that spins are initially uniformly distributed. The propagator, $P(x \mid x_0, t)$, for this system is well-known:
\begin{equation}
    P(x \mid x_0, t) = \frac{1}{L} + \sum_{m=1}^{\infty} \psi_m(x)\psi_m(x_0)e^{-\lambda_mt},
\end{equation}
where
\begin{equation}
    \psi_m(x) = \sqrt{\frac{2}{L}}\cos{\left(\frac{m\pi x}{L}\right)}, \quad \lambda_m = D\left(\frac{m\pi}{L}\right)^2,
\end{equation}
are the eigenfunctions and eigenvalues (scaled by $D$) of the Laplacian with these boundary conditions. This is the starting part for many models of restrictive microstructure. 

\subsection{Second moment}
The result for $S^{(2)}$ is known and was found by Neuman. \cite{Neuman1974} We will re-derive said result for clarity. We aim to find $C(t_1, t_2)$. As an intermediate step, let us find the \emph{position} correlator, denoted $X(t_1, t_2) = \langle x(t_1) x(t_2) \rangle$. In terms of propagators,
\begin{equation*}
\begin{aligned}
    X(t_1, t_2) = \frac{1}{L}\iiint_{[0,L]^3} &dx_0 dx_1 dx_2\, (x_1 x_2) \\ & \times P(x_1 \mid x_0, t_1) P(x_2 \mid x_1, |t_2 - t_1|),
\end{aligned}
\end{equation*}
using $P(x_0) = 1/L$, and where the absolute time difference $|t_2 - t_1|$ may be used by stationarity. The propagator product in the integrand expands as
\begin{equation*}
    \left(\frac{1}{L} + \psi_a(x_0)\psi_a(x_1)e^{-\lambda_a t_1}\right)\left(\frac{1}{L} + \psi_b(x_1)\psi_b(x_2)e^{-\lambda_b|t_2 -t_1|}\right).
\end{equation*}
Because $\int_0^L \psi_a(x_0) dx_0 = 0$, several terms vanish upon further expansion and regrouping. Ultimately, one is left with an eigenfunction in the time difference, indexed by $m$:
\begin{equation*}
    X(t_1, t_2) =\frac{L^2}{4}+\sum_{m=1}^{\infty}\frac{e^{-\lambda_m|t_2-t_1|}}{L}\left(\int_0^L x \psi_m(x) dx\right)^2,
\end{equation*}
where the constant is from $L^{-2}\int_0^L\int_0^L x_1x_2 dx_1dx_2 = L^2/4$. For the bracketed integral above, one finds that
\begin{equation}\label{eq: base eigen integral restriction}
    \int_0^L x \psi_m(x) dx = -\frac{(2\sqrt{2}) L^{3/2}}{\pi^2}\sum_{m\ge 1,\;\text{odd}}^{\infty}\frac{1}{m^2},
\end{equation}
and $0$ for even $m$, obtained by noting the indefinite result $\int x\cos(ux)dx = \cos(ux)/u^2+x\sin(ux)/u + c$, in which the sine term vanishes due to the limits of integration. Thus,
\begin{equation}\label{eq: bounded pos corr}
    X(t_1, t_2) = \frac{L^2}{4}+\frac{8L^2}{\pi^4}\sum_{m\ge1,\;\text{odd}}^{\infty} \frac{e^{-\lambda_m|t_2-t_1|}}{m^4}.
\end{equation}

Next, consider that the 2-point correlator can be expanded as $C(t_1,t_2) = \langle (x_1 - x_0)(x_2-x_0)\rangle$:
\begin{equation*}
    C(t_1,t_2) = X(0,0) + X(t_1, t_2) - X(0, t_1) - X(0, t_2).
\end{equation*}
Solving for the individual terms, one obtains:
\begin{equation}\label{eq: restriction 2-point}
    C(t_1,t_2) = \frac{L^2}{12}+\frac{8L^2}{\pi^4}\sum_{m\ge 1,\;\text{odd}}^{\infty} \frac{e^{-\lambda_m|t_2-t_1|} - e^{-\lambda_m t_1} - e^{-\lambda_m t_2}}{m^4}.
\end{equation}
In calculating the phase variance --- see again Eq. \eqref{eq: phi2} --- only the $e^{-\lambda_m|t_2 - t_1|}$ term in $C(t_1,t_2)$ will contribute. This is because a separable gradient factor such as $\int_0^T G(t_1) dt_1 = 0$ can be pulled out from all other terms. Terms that do not contain all times (here, $t_1$ and $t_2$) vanish by the echo condition. We refer to terms containing all times as being ``fully-connected.''

Evaluating $I = \int_0^T\int_0^{t_2}e^{-\lambda_m (t_2- t_1)}dt_1dt_2$ piecewise for $t_2\ge t_1$:
\begin{equation*}
    I =  \begin{cases}
        \frac{T}{\lambda_m} -\frac{2}{\lambda_m^2}\left(1-e^{-\lambda_mT/2}\right)  , & t_1, t_2 \in \left[0, \frac{T}{2}\right] \\
         \frac{1}{\lambda_m^2}\left(2e^{-\lambda_mT/2} - e^{-\lambda_mT} - 1 \right) , & t_1 \in \left[0,\frac{T}{2}\right], t_2\in\left[\frac{T}{2}, T\right]  
    \end{cases}.
\end{equation*}
By symmetry of the absolute value, the other sub-domains $t_1,t_2\in\left[\tfrac{T}{2},T\right]$ and $t_1\in\left[\tfrac{T}{2},T\right],t_2 \in\left[0,\tfrac{T}{2}\right]$ yield the same result as summing the above, and one obtains after rearrangement:
\begin{equation}\label{eq: phi2 restriction}
    \langle \phi^2 \rangle = \frac{16\gamma^2g^2 L^4}{D\pi^6}\sum_{m\ge1,\;\text{odd}}^\infty \frac{T - \lambda_m^{-1}\left(3 - 4e^{-\lambda_m T/2} + e^{-\lambda_m T}\right)}{m^6},
\end{equation}
which matches the result of Neuman \cite{Neuman1974} upon substitution into $\ln S^{(2)}=-\langle \phi^2 \rangle/2$. It can be shown by Taylor expansion of the exponential terms to third order that $\lim_{L\rightarrow \infty}\langle\phi^2\rangle = \gamma^2g^2T^3D/6$. This is merely the free diffusion result in Eq. \eqref{eq: decay free}, consistent with an infinitely large domain. 

\subsection{Fourth moment}
From $C(t_1, t_2, t_3,t_4) = \langle (x_1-x_0)(x_2-x_0)(x_3-x_0)(x_4 - x_0)\rangle$, we need only to consider the fully-connected terms within $X(t_1, t_2, t_3, t_4)$ to find $\langle \phi^4 \rangle$, as discussed previously. Let us first assume a specific time ordering, $t_1 \le t_2 \le t_3 \le t_4$:
\begin{multline*}
    X(t_1, t_2, t_3, t_4) = \frac{1}{L}\iiiint_{[0,L]^4} dx_1 dx_2 dx_3 dx_4 (x_1x_2x_3x_4) \\ \times
    P(x_2 \mid x_1, t_2 - t_1) P(x_3 \mid x_2, t_3 - t_2) P(x_4 \mid x_3, t_4 - t_3) ,
\end{multline*}
noting that the integral over $x_0$ is separable and evaluates to $\int_0^L P(x_1 \mid x_0, t_1) dx_0=1/L$. 

There are two fully-connected terms that emerge when the integrand above is expanded by substituting the full form of the propagators. One is the triple product:
\begin{multline*}
    \frac{8L^2}{\pi^4}\sum_{a\ge1,\;\text{odd}}^{\infty}\;\sum_{b=1}^{\infty}\;\sum_{c\ge1,\;\text{odd}}^{\infty}\frac{e^{-\lambda_a(t_2-t_1)}e^{-\lambda_b(t_3-t_2)}e^{-\lambda_c(t_4-t_3)}}{a^2c^2}\\\times \left(\int_0^L x_2\psi_a(x_2)\psi_b(x_2)dx_2\right)
    \left(\int_0^Lx_3\psi_b(x_3)\psi_c(x_3)dx_3\right),
\end{multline*}
where we have inserted the results for $x_1$ and $x_4$, each of which appears only once and thus have the same form shown in Eq. \eqref{eq: base eigen integral restriction}. To solve the bracketed integrals, note that the indefinite result of $\int x \cos(ux)\cos(vx)dx$ contains
\begin{equation*}
    \frac{1}{2}\left[\frac{\cos((u-v)x)}{(u-v)^2}+\frac{\cos((u+v)x)}{(u+v)^2}\right],
\end{equation*}
dropping the sine terms because they vanish by the limits of integration. Thus, one obtains for $x_2$:
\begin{equation*}
    \int_0^L x_2\psi_a(x_2)\psi_b(x_2)dx_2 = - \frac{2L}{\pi^2}\left[\frac{1}{(a-b)^2}+\frac{1}{(a+b)^2}\right],
\end{equation*}
where the index $b$ must be even due to the previous constraint on $a$ being odd, and similarly for $x_3$, but swapping $a$ for $c$. The second term to include is the paired product:
\begin{equation*}
    \frac{64L^4}{\pi^8} \sum_{a\ge1,\;\text{odd}}^{\infty}\;\sum_{c\ge1,\;\text{odd}}^{\infty}\frac{e^{-\lambda_a(t_2 - t_1)}e^{-\lambda_c(t_4 - t_3)}}{a^4c^4},
\end{equation*}
inserting known results. Note that no other paired products are fully-connected. Summing and regrouping, one obtains
\begin{widetext}
\begin{equation}\label{eq: phi4 restriction}
    \langle \phi^4 \rangle =\frac{32L^4}{\pi^8}\sum_{a\ge1,\;\text{odd}}^{\infty}\;\sum_{c\ge1,\;\text{odd}}^{\infty}\left(\, \frac{2I_{a,c}}{a^4c^4} +\sum_{b\ge2,\;\text{even}}^{\infty}\;\frac{I_{a,b,c}}{a^2c^2}\left[\frac{1}{(a-b)^2} + \frac{1}{(a+b)^2}\right]\left[\frac{1}{(b-c)^2} + \frac{1}{(b+c)^2}\right]\right),
\end{equation}
where
\begin{equation}
    I_{a,c} = \sum_{\text{perms}\;t_1,t_2,t_3,t_4}\iiiint_{[0,T]^4} G(t_1)G(t_2)G(t_3)G(t_4)  e^{-\lambda_a(t'_2-t'_1)}e^{-\lambda_c(t'_4-t'_3)}dt'_1dt'_2dt'_3dt'_4,
\end{equation}
and
\begin{equation}
    I_{a,b,c} = \sum_{\text{perms}\;t_1,t_2,t_3,t_4}\iiiint_{[0,T]^4} G(t_1)G(t_2)G(t_3)G(t_4) e^{-\lambda_a(t'_2-t'_1)}e^{-\lambda_b(t'_3-t'_2)}e^{-\lambda_c(t'_4-t'_3)}dt'_1dt'_2dt'_3dt'_4.
\end{equation}
\end{widetext}
The sums in $I_{a,c}$ and $I_{a,b,c}$ are over permutations of the labels $\{t_1, t_2, t_3, t_4\}$ mapped to the ordered set: $t'_1 \le t'_2 \le t'_3 \le t'_4$. For example, given $t_4 \le t_3 \le t_2 \le t_1$, one maps $ \{ t'_1 , t'_2 , t'_3 , t'_4\} \to \{ t_4, t_3, t_2, t_1 \}$. As in the trapped-release model, $I_{a,c}$ and $I_{a,b,c}$ can be evaluated piecewise by simultaneous partitioning and ordering. The expanded forms of $I_{a,c}$, $I_{a,b,c}$, and $\langle \phi^4 \rangle$ are long and cumbersome, and are not provided here. Rather, symbolic MATLAB objects are provided as Supplementary Material. See also Appendix \ref{appx: exp trap integrals} for further details and pseudo-code for how to calculate these quantities. 

For the remainder of this section, we present truncations of the series representations of $\langle \phi^4 \rangle$ and $\langle \phi^2 \rangle$ at $a,b,c,m \leq 101$. This number of terms appears to be sufficient for numerical precision in the regime of $L/\sqrt{DT} < 100$, where $\sqrt{DT}$ is the r.m.s displacement over each gradient application $T/2$, and is the natural lengthscale with which to compare $L$. This is also called the diffusion length. \cite{Hurlimann1995} Numerical justification for the truncation is provided later. We note, however, that the convergence for $\langle \phi^4 \rangle$ can become delicate for very large $L/\sqrt{DT} \gg 100$ (data not shown), and may require more terms than used here. 

\begin{figure}
    \centering
    \includegraphics{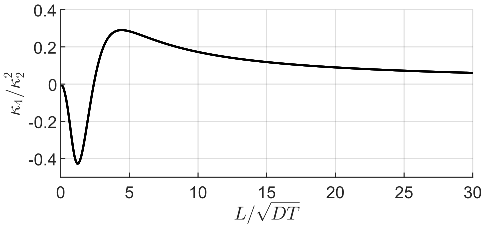}
    \caption{$\kappa_4/\kappa_2^2$ vs. $L/\sqrt{DT}$ for a restricted diffusion in domain length $L$. Parameters $T = 10\;\mathrm{ms}$ and $D = 2\;\mathrm{\mu m^2/ms}$ were fixed, while $L$ was varied. The kurtosis was estimated using $\langle\phi^4 \rangle$ in Eq. \eqref{eq: phi4 restriction} and $\langle \phi^2 \rangle$ in Eq. \eqref{eq: phi2 restriction}, truncated at $a,b,c,m \le 101$.}
    \label{fig: restriction kurt}
\end{figure}

In Fig. \ref{fig: restriction kurt}, $\kappa_4/\kappa_2^2$ is plotted vs. $L/\sqrt{DT}$, varied linearly between $0.1 - 30$ by $0.1$. The parameters $T=10\;\mathrm{ms}$ and $D = 2\;\mathrm{\mu m^2/ms}$ were fixed while $L$ was varied such that the plot can be read from left-to-right as going from a confined to a large domain. There is complicated behavior with multiple peaks. A negative peak can be seen at $L/\sqrt{DT}\approx1.2$, followed by a positive peak at $L/\sqrt{DT} \approx 4.4$, and finally a decay towards $0$ as $L\rightarrow \infty$ and free diffusion is restored. Note again that this behavior is not dependent on $g$.

These regimes of behavior can be thought of as follows. At $L/\sqrt{DT}\ll 1$, $\kappa_4/\kappa_2^2 \approx 0$ follows from a CLT argument. \cite{Hurlimann1995} Then, for appreciable but small $L/\sqrt{DT} \approx 0.5 - 2$, the tails of $P(\phi)$ are effectively truncated by the boundary, leading to a platykurtic shape and negative $\kappa_4/\kappa_2^2$ overall. As the domain grows, $L/\sqrt{DT} \gtrsim 2$, tails are not as strongly affected because many spins will not encounter a wall over the encoding, but spins that begin near walls experience less dephasing due to reflection. This proportion of spins localized near walls leads to central bias in $P(\phi)$ such that it is leptokurtic, with positive $\kappa_4/\kappa_2^2$. The sign change occurs at roughly $L/\sqrt{DT} \approx 2.4$, and the positive peak lies at about $L/\sqrt{DT}\approx 4.4$. In the limit of free diffusion, $L/\sqrt{DT}\rightarrow \infty$, the fraction of spins near walls gradually decreases, and $\kappa_4/\kappa_2^2\rightarrow 0$. 

\begin{figure*}
    \centering
    \includegraphics{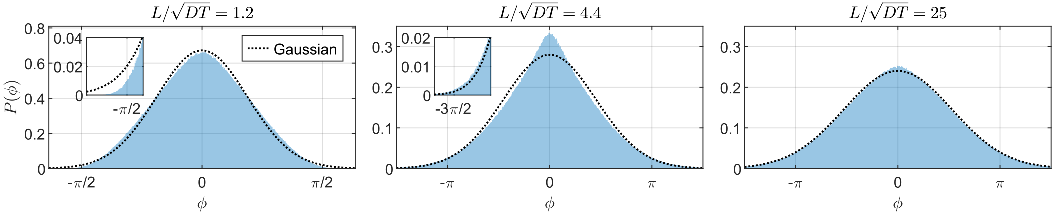}
    \caption{Phase distribution $P(\phi)$ at several values of $L/\sqrt{DT} = [1.2,\,4.4,\,25]$ for fixed $T = 10\;\mathrm{ms}$, $D = 2\;\mathrm{\mu m^2/ms}$, and $g = 0.35\;\mathrm{T/m}$. MC simulated phase (blue) are compared to a Gaussian (dashed), obtained from the mean and standard deviation of simulated data. See main text for simulation methods. From left-to-right, the first plot lies near to the negative peak in Fig. \ref{fig: restriction kurt}, the second to the positive peak, while the third approaches the Gaussian limit. The simulated $\kappa_4/\kappa_2^2$ values are $\approx -0.42,\,0.29, \,0.08$, respectively, in agreement with Fig. \ref{fig: restriction kurt}. Insets highlight behavior at the left tail of the distributions. Note the truncated tails in the first plot vs. approximately Gaussian tails in the second.  }
    \label{fig: restriction phi distr}
\end{figure*}

To illustrate these regimes, examples of $P(\phi)$ are given in Fig. \ref{fig: restriction phi distr} for $L/\sqrt{DT} = [1.2, \, 4.4, \,25]$, with fixed $T = 10\;\mathrm{ms}$, $D = 2\;\mathrm{\mu m^2/ms}$, and $g = 0.35\;\mathrm{T/m}$.
Data were generated by MC simulations with time step $1 \;\mathrm{\mu s}$, step size $\approx 6.3 \times 10^{-2}\;\mathrm{\mu m}$, and $5\times 10^6$ walkers with random and uniformly distributed initial position $\in[0, L]$. Elastic wall collisions were assumed. The regimes discussed above and the rationale for the emergence become clear. There is heavy tail truncation at $L/\sqrt{DT} = 1.2$, then a strong central peak accompanied by nearly Gaussian tails at $L/\sqrt{DT} = 4.4$, and finally a return Gaussianity as $L/\sqrt{DT} = 25$ becomes large, manifesting as a gradual shrinking of the central peak. Importantly, simulated values of $\kappa_4/\kappa_2^2$ match Fig. \ref{fig: restriction kurt}, indicating that negligible bias is introduced by the series truncation at $a,b,c,m \leq 101$. 

\subsection{Predicted signal behavior}

\begin{figure}
    \centering
    \includegraphics{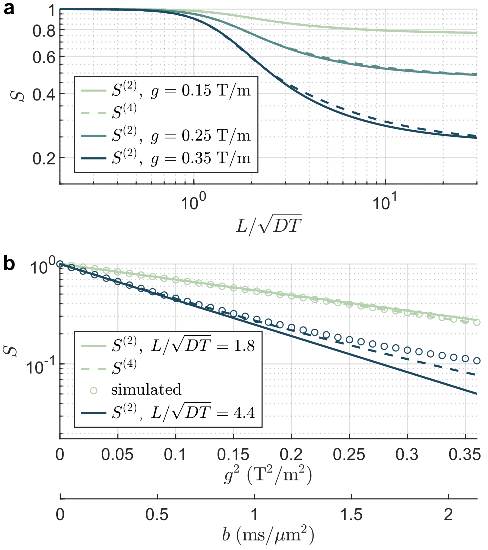}
    \caption{Signal for restriction in domain size $L$. (a) $S^{(2)}$ (solid) and $S^{(4)}$ (dashed) vs. $L/\sqrt{DT}$ for varied $g = [0.15,\, 0.25,\, 0.35]\;\mathrm{T/m}$ (color). Parameters $T = 10\;\mathrm{ms}$ and $D = 2\;\mathrm{\mu m^2/ms}$ were fixed, while $L$ was varied. Truncated approximations were obtained as described in Fig. \ref{fig: restriction kurt} and the main text. (b) $S^{(2)}$ (solid), $S^{(4)}$ (dashed), and MC simulated signals (circles) vs. $g^2$ for fixed values of $L/\sqrt{DT} = [1.8,\,4.4]$ (color), with $g$ varied linearly between $0 - 0.6\;\mathrm{T/m}$ by $0.01$. Other parameters were kept the same as part (a). A secondary axis with $b = \gamma^2 g^2 T^3/12$ is included. Note that the difference between $S^{(2)}$ and $S^{(4)}$ is positive vs. negative for $L/\sqrt{DT} = 1.8$ vs. $4.4$, which correspond to the negative and positive regimes in Fig. \ref{fig: restriction kurt}, respectively.}
    \label{fig: restriction analyt}
\end{figure}

In Fig. \ref{fig: restriction analyt}a, predicted signals $S^{(2)}$ and $S^{(4)}$ are plotted for the same parameters, varying $g$ and $L$. The negative $\kappa_4/\kappa_2^2$ regime ($L/\sqrt{DT} \lesssim 2.4$) leads to a slight overestimate in $S^{(2)}$ compared to $S^{(4)}$, while at larger $L/\sqrt{DT}$, the positive $\kappa_4/\kappa_2^2$ regime results in more notable underestimation in $S^{(2)}$ vs. $S^{(4)}$ --- see the largest $g = 0.35\;\mathrm{T/m}$. Thus, $S^{(2)}$ is an adequate approximation for weak dephasing, but stronger deviations in the direction $S^{(2)} < S^{(4)}$ emerge as $g$ increases. Regarding the ``motional averaging'' regime, \cite{Hurlimann1995} such small-domain signal approximations are surely valid when $L/\sqrt{DT} \ll 1$, but might not extend to $L/\sqrt{DT} \approx 1 - 2$, where the negative kurtosis is appreciable and has a subtle effect on the signal that is more apparent with larger $g$. Interestingly, they may also be applied at $L/\sqrt{DT} \approx 2.3 - 2.5$ (corresponding to the zero-crossing).

In Fig. \ref{fig: restriction analyt}b, $S^{(2)}$ and $S^{(4)}$ are plotted vs. $g^2\propto b$, while fixing $L/\sqrt{DT} = [1.8,\,4.4]$. These values are chosen to lie near the negative and positive peaks shown in Fig. \ref{fig: restriction kurt} (note that $1.8$ was chosen instead of $1.2$ because the former exhibits more signal decay). It can be seen in Fig. \ref{fig: restriction analyt}b that, depending on $L/\sqrt{DT}$, $S^{(4)}$ can be either sub- or super-linear w.r.t. $g^2$, corresponding to positive and negative $\kappa_4/\kappa_2^2$, respectively. The sub-linear behavior in the $L/\sqrt{DT} = 4.4$ case deviates more strongly from $S^{(2)}$ than the super-linear behavior for $L/\sqrt{DT} = 1.8$, which is considerably more subtle. Still, the fact that super-linear decay w.r.t. $g^2$ can emerge for simple restriction may be surprising given its ``anomalous'' \cite{Vilk2022} appearance. Though not seen here for $g^2 < 0.6\;\mathrm{T^2/m^2}$, one may again expect that $S^{(4)}$ eventually exhibits non-physical behavior at large $g$. 

Numerically simulated signals are included in Fig. \ref{fig: restriction analyt}b for comparison. MC methods were the same as the previous subsection but with fewer walkers $5\times 10^5$. The simulation was run once per $g$. Similar to the results shown in Figs. \ref{fig: poisson signal}b and \ref{fig: exp trap release analyt}b, $S^{(4)}$ is reasonably accurate until about $S\lesssim 0.1$, while $S^{(2)}$ is accurate until $S \lesssim 0.7$. Importantly, the simulated signal confirms the slight super-linear behavior for $L/\sqrt{DT} = 1.8$. Note that there remains some deviation for $L/\sqrt{DT} = 4.4$ at large $g^2 \gtrsim 0.25\;\mathrm{T^2/m^2}$, indicating that higher-order cumulants may be significant in this regime.    

\section{Discussion}

\subsection{Summary of findings}

In this manuscript, the exact excess phase kurtosis $\kappa_4/\kappa_2^2$ was derived for the CGSE experiment in several 1D model systems: (1) Poisson pore-hopping, (2) trapped-release with exponentially-distributed release time, and (3) restriction. In each case, $\kappa_4/\kappa_2^2$ was found to depend on a key non-dimensional parameter. In (1), this was the ratio of echo time and mean inter-hop time, $T/\tau_{\text{hop}}$, or mean hop number, and $\kappa_4/\kappa_2^2$ was found to scale linearly with the inverse, $\tau_{\text{hop}}/T$. In (2), this parameter was the product of the release rate and echo time, $\alpha = kT$, and $\kappa_4/\kappa_2^2$ was found to decay roughly log-linearly with $\alpha$. In (3), the parameter was the ratio of domain length and diffusion length, $L/\sqrt{DT}$, and $\kappa_4/\kappa_2^2$ was found to have complicated behavior with multiple distinct regimes: a negative peak was found at $L/\sqrt{DT}\approx 1.2$, and a positive peak at $L/\sqrt{DT} \approx 4.4$. We note that positive kurtosis can become large ($\kappa_4/\kappa_2^2 \gg 1$) for exchange-like behavior such as in models (1) and (2), while it remains relatively bounded ($|\kappa_4/\kappa_2^2|<0.4$) in restriction. We again emphasize that, to our knowledge, these are among the first analytical results for $\kappa_4/\kappa_2^2$ without gross model simplifications such as Gaussian compartments or infinitesimal gradient pulses.

We have also presented signal approximations to second, $\ln S^{(2)} = -\kappa_2/2$, and fourth order, $\ln S^{(4)} = - \kappa_2/2 + \kappa_4/24$. Commensurate with positive phase kurtosis, $\ln S^{(4)}$ was found to decay sub-linearly w.r.t. $g^2\propto b$ in almost all cases, with the notable exception of $L/\sqrt{DT} \approx 1-2$, wherein negative $\kappa_4/\kappa_2^2$ led to slightly super-linear decay. Overall, $S^{(4)}$ was seen to be a good signal approximation for $S \gtrsim 0.1$, while $S^{(2)}$ was a good approximation only for $S \gtrsim 0.7$, or when dephasing is weak --- see again Figs. \ref{fig: poisson signal}b, \ref{fig: exp trap release analyt}b, and \ref{fig: restriction analyt}b. Beyond this weak dephasing regime, the GPA also holds for (1) $T/\tau_{\text{hop}} \gtrsim 10$, (2) $\alpha \gtrsim 5$, and (3) $L/\sqrt{DT} \ll 1$, $L/\sqrt{DT} \approx 2.3 - 2.5$, or $L/\sqrt{DT}\gtrsim 20$ given a heuristic bound of $|\kappa_4/\kappa_2^2| \lesssim 0.1$ --- see again Figs. \ref{fig: poisson signal}a, \ref{fig: trapping release kurt}, and \ref{fig: restriction kurt}. Together, these findings provide clear guidelines for when the GPA may be applied for this experiment and these paradigmatic model systems.  

A more formal justification of the weak dephasing regime is given here. Consider that in the neighborhood of $\phi = 0$, any symmetric $P(\phi)$ is well-approximated by a Gaussian. \cite{Tierney1986} To see this, one can look at the Maclaurin expansion of $\ln P(\phi)$,
\begin{equation*}
    \ln P(\phi) \approx \ln P(0) + \frac{1}{2} \frac{d^2}{d\phi^2} \ln P(\phi) \Bigr|_{\phi = 0} \phi^2 + \mathcal{O}(\phi^4),
\end{equation*}
noting that $\tfrac{d}{d\phi}\ln P(\phi)\big|_{\phi =0} = 0$. For small $\phi$, the error term $\mathcal{O}(\phi^4)$ is negligible. Exponentiating yields a Gaussian form: 
\begin{equation}
    P(\phi) \simeq P(0) \exp\left(-\frac{z}{2} \phi^2\right), \quad \text{for\;small\;} \phi,
\end{equation}
where $z = -\tfrac{d^2}{d\phi^2} \ln P(\phi)\big|_{\phi = 0}$ is the negative log curvature at $\phi = 0$. Thus, weak dephasing with $\phi$ concentrated near 0 is itself a sufficient condition for the GPA. Put another way, the initial log signal decay w.r.t. $g^2$ will always be approximately linear since $\phi \propto g$ when $\phi$ is small, roughly speaking. Note that this holds irrespective of the specific microstructure or motion behavior, and is the only regime in which the GPA is universal. A similar argument could be made for $S^{(4)}$.




\subsection{Comparison to the diffusional kurtosis model}

As discussed in the introduction, a central distinction is that diffusional kurtosis is a signal model that purports to quantify non-Gaussianity in the net displacement distribution, whereas the measured quantity is phase, which depends on entire spin trajectories and is sensitive to finite gradient durations. An additional concern is that the diffusional kurtosis model makes use of a signal expansion in powers of $b$ to yield experimental estimates. As written by Jensen \textit{et al.} as \cite{Jensen2005}:
\begin{equation}\label{eq: diff kurtosis est}
    \ln S(b) \approx -bD_{\text{app}} + \frac{1}{6}b^2D^2_{\text{app}}K_{\text{app}} + \mathcal{O}(b^3),
\end{equation}
where $D_{\text{app}}$ and $K_{\text{app}}$ represent apparent diffusivity and excess displacement kurtosis, respectively. 

While the parameters are acknowledged to be apparent, the expansion itself remains problematic. It implies that the $n^{\text{th}}$-order signal terms scale as $(bD)^n$, which corresponds to $(\gamma g T^3 D)^n$ for the CGSE. This does not generally hold, and explicit counterexamples were presented here. In the Poisson pore-hopping model, $\kappa_n \propto T^{n+1} D_{\text{eff}}^n/\tau_{\text{hop}}^{n-1}$, from Eqs. \eqref{eq: Deff poisson} and \eqref{eq: nth cumulant poisson}. In restriction, the leading scaling is $\kappa_2 \propto L^4/D$, see again Eq. \eqref{eq: phi2 restriction} and Neuman. \cite{Neuman1974} Thus, while the diffusional kurtosis model may provide a better fit than GPA-dependent models by including a term $\propto g^4$, the model may, at least in some cases, imply incorrect scaling behavior. N.B., Jensen \textit{et al.} \cite{Jensen2005} state that timing parameters should be fixed and only $g$ varied when estimating $K_{\text{app}}$, which may mask these scaling issues. Still, the problem persists if timings vary between protocols or studies. These discrepancies underscore the fact that phase statistics --- rather than $b$-value based expansions --- are the appropriate way to analyze non-Gaussian signal behavior. We will return to the topic of diffusional kurtosis later when we discuss the estimation of $\kappa_4/\kappa_2^2$ from experiments. 

\subsection{Implications for microstructure signal modeling}

Let us consider what these results imply practically for the broader field of diffusion microstructural MR. Though it has long been known that the signal behavior can be non-Gaussian (indeed, Stejskal \cite{Stejskal1965b} remarked on the possibility in the year that PGSE was introduced \cite{Stejskal1965}), non-Gaussian phase behavior has often been ignored --- as in models based on apparent diffusion coefficient(s) or tensor(s) --- or explained by some form of phenomenological decomposition, i.e., by additively extending second-order models, such as including multiple Gaussian compartments. \cite{Yablonskiy2003} Gaussian decomposition is especially convenient as it corresponds to simple, multi-exponential signal decay. More complicated but nonetheless second-order decompositions use signal approximations in the motional averaging regime, \cite{Neuman1974, Hurlimann1995} wherein Neuman's results for simple geometries, such as Eq. \eqref{eq: phi2 restriction}, are truncated to leading terms. For the CGSE in 1D, the approximation for diffusion between parallel planes is 
\begin{equation}
    \ln S \simeq -\frac{\gamma^2g^2 L^4}{120 D}\left(T - \frac{17}{56}\frac{L^2}{D}\right), \quad L/\sqrt{DT} \rightarrow 0.
\end{equation}
For example, CHARMED, \cite{Assaf2005} AxCaliber, \cite{Assaf2008}, ActiveAx \cite{Dyrby2012}, and NODDI \cite{Zhang2012} all use a similar approximation to model axons as impermeable cylinders, and SANDI \cite{Palombo2020} uses one to model cell bodies or soma as spheres; they do so as part of a larger, multi-compartment signal model.

Here, we have shown that non-Gaussian signal behavior can arise solely from the complicated evolution of $P(\phi)$, even in systems with just one governing length- or timescale. Thus, such signal behavior is not itself an adequate justification for multiple compartments, and one must be careful in making this supposition. Many distinct motion phenomena can result in sublinear signal behavior w.r.t. $g^2$ or $b$, i.e., these effects are degenerate in typical experiments with a single diffusion encoding. \cite{Coelho2019} That aside, models that use a motional averaging approximation may have biased size estimates due to neglecting the effect(s) of phase kurtosis. If measured in the negative $\kappa_4/\kappa_2^2$ regime of $L/\sqrt{DT} \approx 1- 2$, estimates may be inflated due to super-linear decay w.r.t. $g^2$. In the positive regime, size may be underestimated, particularly near the peak at $L/\sqrt{DT} \approx 4.4$. As discussed, the latter is likely to have a stronger effect on the signal. The same concerns may apply to TDS \cite{Parsons2005} and its estimation of the dispersive velocity autocorrelation function. Other authors have also commented on the deficiencies of second-order models and the decompositions based on them, and we refer the reader to refs. \citenum{Novikov2018b, Grebenkov2007, Grebenkov2010, LeBihan2012} for further perspectives.

\subsection{Limitations}

There are several limitations to this work related mainly to its narrow scope. Firstly, dimensionality may play a role in the severity of non-Gaussian effects, and only 1D systems were studied here. Consider that in 2- or 3D, the number of paths that spins may take increases, which could contribute to the averaging process that restores Gaussianity via the CLT --- i.e., one may speculate that $\kappa_4/\kappa_2^2$ generally decreases with dimensionality. This is not a rigorous argument, however, and such behavior likely depends on the specific microstructure and whether it behaves as an ``effective medium'' with macroscopic averaging. \cite{Novikov2010} There is numerical evidence of this trend with dimensionality for CGSE and PGSE experiments in the case of restriction. Balinov \textit{et al.} \cite{Balinov1993} found that spheres usually, but not always, exhibited less deviation from the GPA than parallel planes. Sukstanskii \& Yablonskiy \cite{Sukstanskii2002} expanded on these results by considering planes, cylinders, and spheres at various levels of signal dephasing, with the same general finding that the 1D case usually had the strongest non-Gaussian effects. Extending our findings to 2- and 3D dimensions is a topic for future work. For now, we note that the relationship to the 1D case is likely to be non-trivial given previous findings of exceptions to dimensionality reducing $\kappa_4/\kappa_2^2$. It should be noted, however, that the Poisson pore-hopping and trapped-release models are readily extended to higher dimensions. 

Secondly, only the CGSE experiment was considered here for the trapped-release and restriction models. PGSE is a much more common experiment, and it is known that the two can exhibit drastically different signal behavior even at the same diffusion-weighting or $b$-value. For example, diffusion-diffraction patterns, \cite{Callaghan1991} are a phenomenon unique to PGSE. Though speculative, we do not expect that PGSE sequences will reduce $\kappa_4/\kappa_2^2$ in general. Recall that we found that PGSE had greater kurtosis than CGSE by $\approx 11\%$ in the Poisson-pore hopping model. Additionally, studies have shown that PGSE signals exhibit non-Gaussian power law behavior $S \propto g^{-4}$ in porous media at large $g$, \cite{Sen1995, Frhlich2006} suggestive of the action of large higher-order phase cumulants. The abscissa of diffraction patterns \cite{Callaghan1991, Topgaard2025} may also suggest oscillation and peaks in $\kappa_4/\kappa_2^2$. There may be a rich range of PGSE behaviors not explored in this manuscript, and we stress that intuition for the CGSE experiment does not necessarily translate to PGSE or other types of experiments. Possible extensions to PGSE and TDS are discussed in the next subsection.   

A more fundamental limitation of this work arises from the cumulant expansion itself. A truncated signal approximation such as $\ln S^{(4)}$ is fundamentally a perturbative expansion, and will exhibit non-physical, divergent behavior at large enough $g$, as noted throughout the manuscript. Adding terms could extend the range of $g$ for which a truncated approximation is accurate, but obtaining these terms would be difficult to the point of infeasibility except in special cases, as evidenced by our tedious calculation of $\langle \phi^4 \rangle$ for restriction. We speculate that, at present, no realistically obtainable number of terms can accurately replicate the known asymptotic $-\ln S \propto g^{2/3}$ behavior of the localization regime. \cite{Stoller1991} 

As a result, we must draw a clear distinction between this work and those that aim to solve the Bloch-Torrey equation \cite{Torrey1956} exactly, e.g., the seminal works on the localization regime \cite{Stoller1991, deSwiet1996}, and later developments. \cite{Grebenkov2014, Herberthson2017, Grebenkov2018} It is only by these or similar non-perturbative approaches \cite{Afzali2022} that signal behavior at large gradients can be fully understood. Nonetheless, our elucidation of the excess phase kurtosis represents an important step towards achieving a more complete understanding of the diffusion MR signal. Notably, our findings dispel the notion that the GPA is generally valid, and provide novel intuition on the intermediate regime that lies between the GPA and the localization regime.  

\subsection{Beyond CGSE experiments}

Solving for $\kappa_4/\kappa_2^2$ in PGSE is a promising topic for future work, and we remark briefly on how one might proceed. If rise times can be ignored, PGSE introduces a third time interval where $G(t) = 0$. The same approach of piecewise integration may be applied to find $\langle \phi^4 \rangle$ --- see again Appendix \ref{appx: exp trap integrals} --- but the $4$ time indices are assigned to $3$ rather than $2$ intervals. However,  assignment to the zero interval causes that term to vanish, such that there are effectively the same number of non-zero terms as CGSE, but with the second time interval shifted by the time between gradient pulses. Thus, PGSE should be no more difficult to solve than CGSE from the perspective of computational complexity. 

Another class of experiments to consider is the oscillating or modulated gradients of TDS. \cite{Parsons2005} In TDS, multiple echoes are formed such that the additive property of cumulants (i.e., for independent processes) becomes relevant. If the phase accrued during each echo-forming encoding block is weakly correlated, then the overall phase $\Phi$ is approximately the sum of individual block phases $\phi_j$ over $J$ blocks : $\Phi \approx \sum_{j = 1}^J \phi_j$. This implies that the overall $n^{\text{th}}$ cumulant is $\approx J\kappa_n$, where $\kappa_n$ is the cumulant for a single block. As a result, $\kappa_4[\Phi]/\kappa_2^2[\Phi] \approx (\kappa_4/\kappa_2^2)/J$. The excess phase kurtosis should be inversely proportional to the number of repeated encodings, at least in the ideal case that there are no strong phase correlations between encoding blocks. This trend of decreased kurtosis with an increased number of gradient waveform periods was demonstrated numerically in a study of restricted cylinders by Topgaard, \cite{Topgaard2025} and the above argument is a plausible explanation for those findings. 

For this class of experiment, it is more feasible to use a large number of repeated encodings such that assuming the GPA is reasonable, rather than attempting to treat analytically. This is because the number of non-zero gradient intervals can quickly become large, greatly complicating the calculation of $\langle \phi^4 \rangle$ by piecewise integration. Consider, for example, that even in the simplest case of one waveform period with 4 intervals $\pm g$, the number of possible assignments becomes $7$ choose $3 = 35$, which is a $7$-fold increase in the number of terms to consider than in CGSE or PGSE (note, this grows in a binomial sense with the number of encoding blocks, $\sim 2^{J}$). The rate at which $P(\phi)$ becomes Gaussian in different geometries, dimensions, and gradient waveform shapes, however, is not yet known. The speculated linear approach is only an approximation that, again, may fail when phase is correlated between encoding blocks. One such scenario may be when the waveform period is close to the ratio of structure size and diffusion length over said period.         

\subsection{Estimating the excess phase kurtosis}

While the message of this manuscript has been cautionary to this point, the excess phase kurtosis is also a measurable and potentially informative quantity. As $\kappa_4/\kappa_2^2$ is unit-less, the signal w.r.t. $g$ for some fixed $T$ can be estimated using arbitrary polynomial coefficients:
\begin{equation}
    \ln S (g) \simeq -c_1 g^2 + c_2 g^4 + \mathcal{O}(g^6), \quad \text{for\;small} \;g
\end{equation}
such that
\begin{equation}
     \frac{\kappa_4}{\kappa_2^2} \approx 6 \frac{c_2}{c_1^2},
\end{equation}
taking into account the cumulant expansion prefactors. This should be a good approximation provided it is estimated in the regime where $S^{(4)}$ is accurate. As discussed previously, $S \gtrsim 0.1$ is a heuristic. A principled check would be to include the next-order term, $-c_3 g^6$, in a separate fit to verify that
\begin{equation*}
    \left|\frac{c_3 g^6}{120}\right| \ll \left|\frac{c_2 g^4}{24}\right|
\end{equation*}
for the largest $g$ considered. This is in essence a recasting of Eq. \eqref{eq: diff kurtosis est}, but which makes no assumptions about the particular form of the coefficients w.r.t. powers of $g$, and thus avoids the pitfalls associated with using the $b$-value and any incorrect scalings that might be implied. 

It should be stated, however, that if the estimation of $K_\text{app}$ is carried out with fixed timing parameters and only varying $g$, as originally suggested, \cite{Jensen2005} the estimates should yield the same value, i.e., $K_{\text{app}} \equiv \kappa_4/\kappa_2^2$. The discrepancies between what we propose here and the diffusional kurtosis model thus lie in their assumptions and interpretation --- it is not necessary to invoke infinitesimal gradients and the bijectivity of phase and displacement to identify the scaling behavior of $\ln S$ w.r.t. $g$. Indeed, the reason that $K_{\text{app}}$ is apparent and different from the kurtosis of the net displacement distribution is the fact that gradients have finite duration. We posit that what has been measured in diffusional kurtosis works is more precisely the excess phase kurtosis, and it should be acknowledged as such without invoking displacement statistics, except in cases where gradients are truly short (relative to diffusion times). Put another way, the diffusional kurtosis model is likely to be operationally valid despite flawed theoretical underpinning; for finite gradients, $K_{\text{app}}$ is not what it is purported to be, and represents a phase, rather than a displacement quantity.

That being said, a single estimate of $\kappa_4/\kappa_2^2$ may not be very interesting, as a number of different phenomena can yield the same value. It may be more informative to vary $T$ and then repeat the estimation to yield $\kappa_4/\kappa_2^2$ w.r.t. $T$, similar to Figs. \ref{fig: poisson signal}b, \ref{fig: trapping release kurt}, and \ref{fig: restriction kurt}. The  range of $g$ values can be adjusted to stay in the valid signal regime $S\gtrsim 0.1$, i.e., by decreasing $g$ as $T$ increases. Qualitatively, according to the model systems studied here, a linear decay in $\kappa_4/\kappa_2^2$ w.r.t. $T$ might imply pore-hopping behavior. A roughly log-linear decay would be consistent with first-order release or barrier-limited exchange. The presence of peaks would imply restriction, with the peak(s) location depending on size. Looking beyond, a combination of restriction and exchange might attenuate and shift these peaks. A constant kurtosis would be consistent with (non-exchanging) Gaussian compartments, \cite{Novikov2018b} and so forth. Recently, a similar protocol which measured $K_{\text{app}}$ at varied PGSE timings was proposed by Lee \textit{et al.}, \cite{Lee2024} demonstrating the potential of such an approach.







\section*{Conclusion}

We have presented analytical expressions for the excess phase kurtosis in the constant gradient spin echo experiment. Several one-dimensional model systems were studied, including both exchange-like systems and the foundational case of restricted diffusion. Results were validated by Monte Carlo simulations. Our findings shed light on the validity of the common Gaussian phase approximation, and provide clear guidelines for when it may be applied in said experiment. More broadly, we have laid the theoretical groundwork for exploring higher-order phase statistics in other experiments or systems.

\appendix
\section{Calculation of $\langle \phi^4 \rangle$ by piecewise integration}\label{appx: exp trap integrals}

Here we demonstrate how to calculate $\langle \phi^4 \rangle$ by piecewise integration for the models presented in the main text. We first address the trapped-release model and Eq. \eqref{eq: exp trap phi4}. To begin, we consider whether each time index is assigned to the interval $\left[0, \tfrac{T}{2}\right]$ or $\left[\tfrac{T}{2}, T\right]$. The assignment of 4 times to 2 intervals can be done 5 ways. Note that in general, the number of distinct assignments is the binomial coefficient $3+m$ choose $m-1$ for $m$ distinct intervals --- in this case, $5$ choose $1 = 5$. Here, the gradient term is simple, and $G(t_1)G(t_2) G(t_3) G(t_4) = \pm\gamma^4 g^4$: positive for an even parity assignment (e.g., 2-and-2 in each), and negative for odd. The ordering of times (e.g., $t_1 \leq t_2 \leq t_3 \leq t_4$) must also be considered given the form of $I_{[a,b][c,d]}$ and the presence of different time minimums. There are $4! = 24$ orderings per assignment. For example, assignment of the $2$ smaller times into $\left[0, \tfrac{T}{2}\right]$ with ordering $t_1 \leq t_2 \leq t_3 \leq t_4$ results in the limits of integration:
\begin{equation*}
    \int_0^{t_2}dt_1\int_0^{T/2}dt_2\int_{T/2}^{t_4} dt_3\int_{T/2}^{T} dt_4.
\end{equation*}
Given the Wick pair $I_{[1,3],[2,4]}$, one substitutes $\min{(t_a,t_b)} = t_1$, $\min{(t_c,t_d)} = t_2$, and $t_< = t_1$ into the integrand in Eq. \eqref{eq: exp trap wick pair}. This proceeds in a similar fashion for all $5$ assignments $\times$ $24$ orderings $\times$ $3$ Wick pairs $=360$ distinct sub-domain integrals to be summed to yield $\langle \phi^4\rangle$. Note that the orderings and assignments cover the hypercube exactly once.

We proceed via symbolic programming. A pseudo-code outline is given in Algorithm \ref{alg: phi4 trapping}, written in roughly MATLAB-style syntax. We use round parentheses to denote array indexing or function calls, and square brackets to indicate data or symbolic array declarations. In short, the algorithm loops over the Wick pairs, time orderings, and assignments to set up the appropriate sub-domain integral in terms of bounds and the time minimums, then adds the results together. 
\begin{figure}
\begin{algorithm}[H]
\caption{Calculating $\langle \phi^4 \rangle$ for the trapped-release model}\label{alg: phi4 trapping}
    \begin{algorithmic}[1]
    \Require {$t_1, t_2, t_3, t_4 , T, \gamma, g, D, k \ge 0$}
    \State {$\text{times} \gets [t_1, t_2, t_3, t_4]$}
    \State {$\mathcal{W} \gets [[1,2,3,4];\, [1,3,2,4];\, [1,4,2,3]]$} \Comment{Wick pairs, $[a,b],[c,d]$}
    \State {$\mathcal{P} \gets \text{permutations}([1,2,3,4])$} \Comment{$24 \times 4$ matrix} 
    \State {$\phi_4 = [\;]$} \Comment{to store piecewise integrals}
    
    \For {$i = 1:3$}  \Comment{loop over 3 sets of Wick pairs} 
        \State {$a,b, c,d \gets \mathcal{W}(i)$}
        \For {$j = 1:24$}  \Comment{loop over permutations/ordering}
            \State{$t_{\text{ord}} \gets \text{times}(\mathcal{P}(j))$} \Comment{ordered times, e.g., $[t_4,t_3,t_2,t_1]$}
            \State{$t_{\text{min}} \gets t_{\text{ord}}(1)$} \Comment{smallest time}
            \State{$a_{\text{pos}} \gets \text{find}(\mathcal{P}(j) = a)$}
            \State{$b_{\text{pos}} \gets \text{find}(\mathcal{P}(j) = b)$}
            \State{$c_{\text{pos}} \gets \text{find}(\mathcal{P}(j) = c)$}
            \State{$d_{\text{pos}} \gets \text{find}(\mathcal{P}(j) = d)$}
            
            \If {$a_{\text{pos}} < b_{\text{pos}}$}
                \State{$t_{ab} \gets t_{\text{ord}}(a_{\text{pos}})$} \Comment{$t_{ab}$ denotes $\text{min}(t_a, t_b)$}
            \Else
                \State{$t_{ab} \gets t_{\text{ord}}(b_{\text{pos}})$}
            \EndIf
            \If {$c_{\text{pos}} < d_{\text{pos}}$}
                \State{$t_{cd} \gets t_{\text{ord}}(c_{\text{pos}})$} \Comment{$t_{cd}$ denotes $\text{min}(t_c, t_d)$}
            \Else
                \State{$t_{cd} \gets t_{\text{ord}}(d_{\text{pos}})$}
            \EndIf
            \For{$m = 0:4$}    \Comment{loop over assignments}
                            \\ \Comment{$m$ denotes the no. of times $\in [0, T/2]$}
                \For{$n = 1:4$} \Comment{find integral bounds}
                    \If {$n < m$}
                        \State{$lb(n) \gets 0$} \Comment{integral lower bound}
                        \State{$ub(n) \gets t_{\text{ord}}(n + 1)$} \Comment{integral upper bound}
                    \Else
                        \If {$n = m$}
                            \State{$lb(n) \gets 0$}
                            \State{$ub(n) \gets T/2$}
                        \Else
                            \State{$lb(n) \gets T/2$}
                            \If {$n = 4$}
                                \State{$ub(n) \gets T$}
                            \Else
                                \State{$ub(n) \gets t_{\text{ord}}(n + 1)$}
                            \EndIf
                        \EndIf
                    \EndIf
                \EndFor
                \If {$\text{mod}(m, 2) = 1$}
                    \State{$g_{\text{sign}} = -1$} \Comment{sign of gradient terms}
                \Else
                    \State{$g_{\text{sign}} = 1$}
                \EndIf
                \State{$\mathcal{I} \gets \text{plug}\; (k, D,t_{\text{min}}, t_{ab}, t_{cd})\;\text{into\;Eq. 36}$}
                \Comment{integrand}
                \State{$\mathcal{I} \gets g_{\text{sign}}\times \gamma^4 g^4\times \mathcal{I}$}
                \State{$I_1 \gets \text{integrate}(\mathcal{I},\, t_{\text{ord}}(1),\, lb(1),\, ub(1)),$} 
                \State{$I_2 \gets \text{integrate}(I_1,\, t_{\text{ord}}(2),\, lb(2),\, ub(2)),$}
                \State{$I_3 \gets \text{integrate}(I_2,\, t_{\text{ord}}(3),\, lb(3),\, ub(3)),$}
                \State{$I_4 \gets \text{integrate}(I_3,\, t_{\text{ord}}(4),\, lb(4),\, ub(4)),$}
                \State{$\phi_4 \gets \phi_4 + I_4$} \Comment{sum to result}
            \EndFor
        \EndFor
    \EndFor
    \end{algorithmic}
\end{algorithm}
\end{figure}
Let us go through an example piecewise integral as well. Continuing with the example mentioned above, we have from Eq. \eqref{eq: exp trap wick pair} that the contribution to $\langle \phi^4\rangle$ for this particular Wick pair, time ordering, and assignment is:
 \begin{multline*}
             D^2kT^2\int_0^{T/2}\int_0^{t_2}  t_1t_2\left(\frac{1-e^{-kt_1}}{k}\right) \\ -(t_1 + t_2) \left[\frac{1 - e^{-kt_1}(1+kt_1)}{k^2}\right] \\- \frac{2 + e^{-kt_1}(k^2t^2_1+2kt_1+2)}{k^3}\,dt_1dt_2,
 \end{multline*}
noting that $t_3$ and $t_4$ do not appear in the integrand and thus yield $\int_{T/2}^T\int_{T/2}^T dt_3dt_4= T^2/4$. This evaluates to 
\begin{equation*}
    \frac{D^2T^3}{128k^4}\left(\alpha^4 - 8\alpha^3 - 16\alpha^2 - 192 \alpha + 384 e^{-\alpha/2} + 384\right),
\end{equation*}
where again $\alpha = kT$, with $k = 1/\langle\tau_{\text{rel}} \rangle$. 

Let us now consider the restriction model. Much of the algorithm will be similar in how it handles the integration bounds, general loop structure (without the Wick pairs), and gradient terms. We need only to adjust the integrand and also introduce the eigensum structure. These changes are described in Algorithm \ref{alg: phi4 restr}, where unless otherwise stated, the variables are the same as in Algorithm \ref{alg: phi4 trapping}. Note that the $a = c$ and $a \neq c$ cases need to be handled separately --- see again Eq. \ref{eq: phi4 restriction}. In the first part of Algorithm \ref{alg: phi4 restr}, the integral terms are solved symbolically in terms of eigenvalues $\lambda_a, \lambda_b, \lambda_c$, cased separately for when the indices $a$ and $c$ are equal or not equal. The final loop calculates $\langle \phi^4 \rangle$ up to a specified number of terms. MATLAB objects for $I_{ac,\,a\neq c}$, $I_{ac,\,a= c}$, $I_{abc,\,a\neq c}$, and $I_{abc,\,a = c}$ as described in the Algorithm are provided as Supplementary Material. The file names are `phi4\_a\_neq\_c\_term2.mat', `phi4\_a\_eq\_c\_term2.mat', `phi4\_a\_neq\_c.mat', and `phi4\_a\_eq\_c.mat', respectively.   
\begin{figure}
\begin{algorithm}[H]
\caption{Modifications to Algorithm 1 for restriction}\label{alg: phi4 restr}
    \begin{algorithmic}[1]
    \Require{$t_1, t_2, t_3, t_4, T, \gamma, g, D, L \ge 0$}
    \Require{$a,b,c\;\text{are\;positive integers}$}
    \State{$I_{ac, a\neq c},\,  I_{ac, \,a =c},\, I_{abc, \, a\neq c},\, I_{abc, \, a =c} \gets []$} \Comment{to store results} 
    \For{$i = 1:24$}
        \State{$t_{\text{ord}} \gets \text{times}(\mathcal{P}(i))$} \Comment{ordered times}
        \State{$\tau_1 \gets t_{\text{ord}}(2) - t_{\text{ord}}(1)$} \Comment{time differences in Eqs. 45, 46}
        \State{$\tau_2 \gets t_{\text{ord}}(3) - t_{\text{ord}}(2)$}
        \State{$\tau_3 \gets t_{\text{ord}}(4) - t_{\text{ord}}(3)$}
        \State{$\mathcal{I}_{ac,\, a\neq c} \gets e^{-\lambda_a \tau_1} e^{-\lambda_c \tau_3}$} \Comment{one of the integrand terms}
        \State{$\mathcal{I}_{ac,\,a = c} \gets e^{-\lambda_a (\tau_1 + \tau_3)}$} \Comment{case for $a = c$}
        \State{$\mathcal{I}_{abc,\, a\neq c} \gets e^{-\lambda_a \tau_1} e^{-\lambda_b \tau_2} e^{-\lambda_c \tau_3}$}
        \State{$\mathcal{I}_{abc,\,a = c} \gets e^{-\lambda_a (\tau_1 + \tau_3)} e^{-\lambda_b \tau_2}$} 
        \For{$j = 0:4$} \Comment{assignments}
            \State{$\mathcal{I}_{ac,\,a \neq c} \gets g_{\text{sign}} \times \gamma^4 g^4 \times \mathcal{I}_{ac,\,a\neq c}$}
            \State{$I_{1} \gets \text{integrate}(\mathcal{I}_{ac},\, t_{\text{ord}}(1),\, lb(1),\, ub(1))$}
            \State{$\vdots$} \Comment{nested integrals} 
            \State{$I_{4} \gets \text{integrate}(I_{3,\, ac},\, t_{\text{ord}}(4),\, lb(4),\, ub(4))$}
            \State{$I_{ac,\,a\neq c} \gets I_{ac,\,a\neq c} + I_4$}
            \State{$\vdots$} \Comment{repeat for other integrand terms}
        \EndFor
    \EndFor
\\
    \For{$a = 1:2:M$} \Comment{odd $a$ up to $M$}
        \State{$\lambda'_a = D(a\pi/L)^2$}  \Comment{specific eigenvalue}
        \For{$b = 2:2:M-1$} \Comment{even $b$ up to $M-1$}
            \State{$\lambda'_b = D(b\pi/L)^2$}
            \For{$c = 1:2:M$} 
                \State{$\lambda'_c = D(c\pi/L)^2$} 
                \If {$a = c$} \Comment{substitute specific eigenvalues}
                    \State{$I_{a,c} = \text{sub}(I_{ac,\,a = c},\,\lambda_a,\, \lambda'_a)$} 
                    \State{$I_{a,b,c} = \text{sub}(I_{abc,\,a = c},\,[\lambda_a, \lambda_b],\, [\lambda'_a, \lambda'_b])$}
                \Else
                    \State{$I_{a,c} = \text{sub}(I_{ac,\,a \neq c},\,[\lambda_a, \lambda_c]\, [\lambda'_a,\lambda'_c])$} 
                    \State{$I_{a,b,c} = \text{sub}(I_{abc,\,a \neq c},\,[\lambda_a, \lambda_b, \lambda_c]\, [\lambda'_a, \lambda'_b,\lambda'_c])$} 
                \EndIf
                \State{$\text{term} \gets \text{plug\;in} (L, a,b,c, I_{a,c}, I_{a,b,c})$ to Eq. \eqref{eq: phi4 restriction}}
                \State{$\phi_4 \gets \phi_4 + \text{term}$}
            \EndFor
        \EndFor
    \EndFor
    \end{algorithmic}
\end{algorithm}
\end{figure}

\section*{Acknowledgements}

TXC, NHW, and PJB were supported by the intramural research program (IRP) of the \textit{Eunice Kennedy Shriver} National Institute of Child Health and Human Development (NICHD). NHW was partially funded by the Department of Defense in
the Military Traumatic Brain Injury Initiative (MTBI\textsuperscript{2}) under award HU0001-24-2-0051.

\section*{Contributions}

TXC performed all derivations, simulations, and analysis, and wrote the original draft. All authors contributed to the conceptualization and editing of the manuscript.   

\section*{Declarations}

The authors declare no competing financial interests. Data and the original MATLAB code used for calculations and simulations are available upon reasonable request.

The views, information or content, and conclusions presented do not necessarily represent the official position or policy of, nor should any official endorsement be inferred on the part of, the Uniformed Services University, the Department of Defense, the U.S. Government, or The Henry M. Jackson Foundation for the Advancement of Military Medicine, Inc. The findings and conclusions presented in this paper are those of the author(s) and do not necessarily reflect the views of the NIH or the U.S. Department of Health and Human Services.

\nocite{*}
\section*{References}
\bibliography{refs}

\end{document}